\documentstyle[seceq,epsf,preprint]{ptptex}
\newcommand{\bra}[1]{\langle {#1} |}     %%
\newcommand{\ket}[1]{| {#1} \rangle}     %%
\newcommand{\rket}[1]{| {#1} )}     %%
\newcommand{\rdket}[1]{|\!| {#1} )}     %%
\newcommand{\dket}[1]{|\!| {#1} \rangle}     %%
\newcommand{\lsim}{{\stackrel{<}{\sim}}}
\title{%        %You can use \\ for explicit line-break
Deformed Boson Scheme Stressing Even-Odd \\
Boson Number Difference. I
}
\subtitle{Various Forms of Boson-Pair Coherent State
}    %use this when you want a subtitle

\author{%       %Use \sc for the family name
Atsushi {\sc Kuriyama},$^{1}$ 
Constan\c{c}a {\sc Provid\^encia},$^{2}$ \\
Jo\~ao da {\sc Provid\^encia},$^{2}$ Yasuhiko {\sc Tsue}$^{3}$ 
and Masatoshi {\sc Yamamura}$^{1}$
}

\inst{%         %Affiliation, neglected when [addenda] or [errata]
$^1$Faculty of Engineering, Kansai University, Suita 564-8680, Japan\\
$^{2}$Departamento de Fisica, Universidade de Coimbra, P-3000 Coimbra, 
Portugal\\
$^{3}$Physics Division, Faculty of Science, Kochi University, Kochi 780-8520, 
Japan
}

\recdate{%      %Editorial Office will fill in this.
\today
}

\abst{%       %this abstract is neglected when [addenda] or [errata]
An idea of the deformed boson scheme developed by the present authors is 
applied to the case of boson-pair coherent state. 
In this state, even and odd boson number difference is stressed and various 
concrete examples of the boson-pair coherent states are shown. 
Concerning some of them, mutual relations are investigated in connection 
to the $su(1,1)$-algebra. 
A formulation in terms of the MYT boson mapping is also performed. 
}

\begin{document}

\maketitle

\section{Introduction}

It may be interesting for theoretical studies of many-body systems to 
investigate time evolution of quantal systems described in terms of 
boson operators. To complete this task, the 
time-dependent variational method is 
quite powerful. In this case, the first task is to prepare a trial state 
containing variational parameters. For boson systems, conventional boson 
coherent states may be of the simplest and the most popular form. 
However, the use of the conventional boson coherent states does not always 
give us satisfactory results. For example, recent numerical 
analysis performed by the present authors shows that, compared with certain 
deformed boson coherent states, the conventional one gives us rather poor 
results.\cite{one} 
This numerical analysis has been performed along the proposal of the 
deformed boson scheme by the present authors.\cite{two,three,four,five}
Following the idea by Penson and Solomon,\cite{six} 
our starting idea was presented in Ref.\citen{two}. 
Especially, multiboson coherent state was discussed in 
Refs.\citen{two,three,four}. 
Using the MYT boson mapping\cite{seven} together, the multiboson coherent 
state is formulated in the framework of the deformed boson scheme 
presented in Ref.\citen{two}.

In Ref.\citen{four}, the boson-pair coherent state, that is, the simplest 
multiboson coherent state, is formulated in the framework in which not only 
boson-pairs but also unpaired boson are taken into account. Of course, 
the system adopted there consists of one kind of boson operator 
$({\hat c}, {\hat c}^*)$. Further, as was already mentioned, 
the MYT boson mapping is used and two contrastive boson-pair coherent 
states are discussed. These two boson-pair coherent states were extended 
to the case consisting of two kinds of boson operators.\cite{one} 
Therefore, it may be interesting to investigate other forms of the 
boson-pair coherent states.

Main aim of this paper is to investigate systematically various forms of 
the boson-pair coherent states including two forms discussed in 
Ref.\citen{four}. Further, we formulate our idea in the framework 
of the original boson space. In other words, we adopt the boson mapping. 
Two cases are separately but intimately treated. 
One is the case which consists of even-boson number and the other of 
odd-boson number and, then, the formalism is immediately applicable 
to the Hamiltonian given in terms of ${\hat c}^*{\hat c}$, ${\hat c}^{*4}$, 
${\hat c}^{*3}{\hat c}$, ${\hat c}^{*2}{\hat c}^2$, ${\hat c}^*{\hat c}^3$ 
and ${\hat c}^4$. Therefore, the effect of even-odd boson number difference 
can be investigated. Basic idea can be summarized as follows : 
We decompose conventional boson coherent state expressed in terms of 
the exponential type into parts expressed in terms of the hyperbolic types. 
Then, each part is deformed and we have various deformed boson-pair coherent 
states stressing the even-odd boson number difference. 
In these states, three forms are investigated in rather detail. 
Two states have the forms proposed by the present authors.\cite{four} 
Then, the third form is an interesting state and, in this paper, we discuss 
its outline.

In \S 2, after giving the framework of our basic idea, the conventional 
boson coherent state is treated in our framework. 
In \S 3, the conventional boson coherent state, which is of the exponential 
form, is decomposed into the hyperbolic cosine and sine form. 
Both are expressed as appropriate superpositions of even- and odd-boson 
numbers, respectively, which stress the even-odd boson number difference. 
Section 4 is devoted to giving an idea of the deformed boson scheme for 
the two states introduced in \S 3, and in \S 5 various examples are shown. 
In \S\S 6 and 7, two deformations are discussed. The deformation in \S 6 
leads to the form satisfying the relation of the Poisson bracket 
of the $su(1,1)$-algebra exactly and in \S 7 the form satisfying the relation 
approximately. Finally, the deformations developed in this paper are summarized in the language of the MYT boson mapping.

\section{Framework of the basic idea and its application to 
the conventional boson coherent state}

We are concerned in a many-body system described in terms of 
one kind of boson operator $({\hat c}, {\hat c}^*)$. 
For the time-dependent variational method, the first task is to 
prepare a trial state for the variation, which, in this paper, we denote 
$\ket{tr}$. 
The state $\ket{tr}$ contains a set of complex parameters for the 
variation, $(c, c^*)$. For these parameters, we require the following 
condition :
% for $(c,c^*)$ : 
\begin{eqnarray}\label{2-1}
& &\bra{tr} \partial_c \ket{tr}=c^*/2-i\partial S/\partial c\ , 
\nonumber\\
& &\bra{tr} \partial_{c^*} \ket{tr}=-c/2-i\partial S/\partial c^* \ . 
\end{eqnarray}
Here, $S$ denotes a real function of $(c,c^*)$. 
The condition (\ref{2-1}) was already used in various occasions 
including the TDHF theory in the canonical form, for example, in 
Ref.\citen{eight}. 
If $(c,c^*)$ obeys the condition (\ref{2-1}), $(c,c^*)$ plays a role of 
the canonical variable of classical mechanics in the boson type 
($c=(1/\sqrt{2})(q+ip)\ , \ c^*=(1/\sqrt{2})(q-ip)$). 
In the time-dependent variational method, we must calculate the 
quantity $\bra{tr}i\partial_t \ket{tr}$. 
If $(c,c^*)$ obeys the condition (\ref{2-1}), it can be expressed in the form 
\begin{equation}\label{2-2}
\bra{tr} i\partial_t \ket{tr}=i(1/2)(c^*{\dot c}-{\dot c}^* c)+{\dot S} \ .
\end{equation}
Since ${\dot S}$ denotes the derivative with respect to $t$, this term 
does not give any influence on the time-dependent variation. 
Further, we must calculate 
the expectation value of the Hamiltonian under investigation, 
$\bra{tr} {\hat H} \ket{tr}$. 
Of course, ${\hat H}$ is a function of $({\hat c}, {\hat c}^*)$. 
If we are concerned in boson-pair correlations, the Hamiltonian expressed 
as a function of ${\hat c}^{*2}$, ${\hat c}^2$ and ${\hat c}^*{\hat c}$ must 
be adopted. In other form, ${\hat H}$ is expressed in terms of 
${\hat \tau}_{\pm,0}$ defined as 
\begin{equation}\label{2-3}
{\hat \tau}_+={\hat c}^{*2}/2 \ , \qquad
{\hat \tau}_-={\hat c}^2/2 \ , \qquad
{\hat \tau}_0={\hat c}^*{\hat c}/2 +1/4\ .
\end{equation}
The set $({\hat \tau}_{\pm, 0})$ obeys the $su(1,1)$-algebra : 
\setcounter{equation}{2}
\begin{subequations}
\begin{equation}\label{2-3a}
[{\hat \tau}_+ , {\hat \tau}_- ]=-2{\hat \tau}_0 \ , \qquad
[{\hat \tau}_0 , {\hat \tau}_\pm ]=\pm {\hat \tau}_\pm \ . 
\end{equation}
\end{subequations}
Therefore, it may be indispensable to investigate the boson-pair 
correlations in connection with the $su(1,1)$-algebra. In this paper, 
mainly, we treat the expectation values of 
${\hat \tau}_{\pm,0}$ for $\ket{tr}$. In \S 7, we contact with the expectation 
values of the products of ${\hat \tau}_{\pm,0}$. 
Hereafter, the expectation value of the operator ${\hat O}$ is denoted as  
\begin{equation}\label{2-4}
(O)_{tr}= \bra{tr}{\hat O} \ket{tr} \ .
\end{equation}
Further, as a classical counterpart of $({\hat \tau}_{\pm,0})$, 
we define a set $(\tau_{\pm,0})$ obeying the relation of the Poisson 
bracket 
\begin{equation}\label{2-5}
[ \tau_+ , \tau_- ]_P=(-i)(-2\tau_0) \ , \qquad
[ \tau_0 , \tau_\pm ]_P=(-i)(\pm\tau_\pm) \ . 
\end{equation}
The above is the basic framework of our idea.

The simplest and the most popular example of $\ket{tr}$ may be the 
boson coherent state : 
\begin{equation}\label{2-6}
\ket{tr}=\ket{cr^0}=\left(\sqrt{\Gamma_{cr}^0}\right)^{-1}\dket{ex^0} \ , 
\qquad (\bra{cr^0}cr^0 \rangle =1) 
\end{equation}
\vspace{-0.7cm}
\setcounter{equation}{5}
\begin{subequations}
\begin{eqnarray}
& &\dket{ex^0}=\exp (\gamma {\hat c}^*)\ket{0} \ , \quad ({\hat c}\ket{0}=0) 
\qquad\qquad\qquad
\label{2-6a}\\
& &\Gamma_{cr}^0=\exp(|\gamma|^2) \ . 
\label{2-6b}
\end{eqnarray}
\end{subequations}
Here, $(\gamma, \gamma^*)$ denotes a set of complex parameters which 
should be expressed in terms of $(c,c^*)$. The state $\ket{cr^0}$ 
satisfies the relation 
%\begin{subequations}\label{2-7}
\begin{equation}\label{2-7}
{\hat c}\ket{cr^0}=\gamma\ket{cr^0} \ , \quad 
\hbox{\rm i.e.,}\quad {\hat \tau}_-\ket{cr^0}=(\gamma^2/2)\ket{cr^0} \ . 
\end{equation}
Then, $(\tau_{\pm,0})_{cr^0}$, the expectation values of 
$({\hat \tau}_{\pm,0})$ for $\ket{cr^0}$, are given in the following set : 
\begin{equation}\label{2-8}
(\tau_+)_{cr^0}=\gamma^{*2}/2 \ , \qquad
(\tau_-)_{cr^0}=\gamma^2/2 \ , \qquad
(\tau_0)_{cr^0}=|\gamma|^2/2+1/4 \ . 
\end{equation}
%\end{subequations}
In the present case, the condition (\ref{2-1}) is reduced to 
\begin{eqnarray}\label{2-9}
& &(1/2)(\gamma^* \partial \gamma/\partial c
-\gamma \partial \gamma^*/\partial c)=c^*/2-i\partial S/\partial c \ , 
\nonumber\\
& &(1/2)(\gamma^* \partial \gamma/\partial c^*
-\gamma \partial \gamma^*/\partial c^*)=-c/2-i\partial S/\partial c^* \ . 
\end{eqnarray}
A possible solution of the relation (\ref{2-9}) is given as 
\begin{equation}\label{2-10}
S=0 \ , \qquad \gamma=c \ , \qquad \gamma^*=c^* \ . 
\end{equation}
Then, we have 
\begin{equation}\label{2-11}
(\tau_+)_{cr^0}=c^{*2}/2 \ , \qquad
(\tau_-)_{cr^0}=c^2/2 \ , \qquad
(\tau_0)_{cr^0}=|c|^2/2+1/4 \ . 
\end{equation}
The form (\ref{2-11}) has two characteristic points. One is as follows : 
The form (\ref{2-11}) satisfies the relation of the Poisson bracket 
(\ref{2-5}) approximately, especially at the large value of $|c|^2$. 
The other is related to the products of ${\hat \tau}_{\pm,0}$ in 
normal order for $({\hat c} , {\hat c}^*)$. 
The expectation values of the products of $({\hat \tau}_{\pm,0})$ are 
exactly equal to the corresponding products of $(\tau_{\pm,0})_{cr^0}$. 
This fact comes from the relation (\ref{2-7}).

As was already mentioned in \S 1, the present authors have investigated 
various variations of the above-treated boson coherent state in terms of 
the deformed boson scheme presented in Ref.\citen{two}. 
However, in these framework, the state 
$\ket{tr}$ is expressed as a superposition of the states not only 
with even- but also with odd-boson numbers. 
Therefore, it may be impossible to investigate effects of even-odd boson 
number difference which may be induced by the boson-pair correlations. 
Of course, in this case, the investigation is limited to two 
cases.\cite{two,three,four}

\section{Two possible trial states derived from the conventional boson 
coherent state}

Our next task is to investigate two forms of the trial states 
derived from the conventional boson coherent state. 
First, we note the following decomposition : 
\begin{equation}\label{3-1}
\dket{ex^0}=\dket{ch^0}+\dket{sh^0} \ , \qquad\qquad\qquad\quad\ 
\end{equation}
\setcounter{equation}{0}
\vspace{-0.8cm}
\begin{subequations}
\begin{eqnarray}
\dket{ch^0}&=&(1/2)\left[\exp(\gamma{\hat c}^*)+\exp(-\gamma{\hat c}^*)\right]
\ket{0} \nonumber\\
&=&\cosh (\gamma{\hat c}^*)\ket{0} \ , 
\label{3-1a}\\
\dket{sh^0}&=&(1/2)\left[\exp(\gamma{\hat c}^*)-\exp(-\gamma{\hat c}^*)\right]
\ket{0} \nonumber\\
&=&\sinh (\gamma{\hat c}^*)\ket{0} \ . 
\label{3-1b}
\end{eqnarray}
\end{subequations}
As is clear from their forms, $\dket{ch^0}$ and $\dket{sh^0}$ consist of 
hyperbolic type superpositions of the even- and the odd-boson number states, 
respectively. Then, the above decomposition suggests us to the following two 
trial states for the variation : 
\begin{subequations}\label{3-2}
\begin{eqnarray}
& &\ket{tr}=\ket{ch^0}=\left(\sqrt{\Gamma_{ch^0}}\right)^{-1}\dket{ch^0} \ , 
\quad (\bra{ch^0}ch^0\rangle =1) 
\label{3-2a}\\
& &\ket{tr}=\ket{sh^0}=\left(\sqrt{\Gamma_{sh^0}}\right)^{-1}\dket{sh^0} \ , 
\quad (\bra{sh^0}sh^0\rangle =1) 
\label{3-2b}
\end{eqnarray}
\end{subequations}
\vspace{-0.8cm}
\begin{subequations}\label{3-3}
\begin{eqnarray}
& &\Gamma_{ch^0}=\cosh |\gamma|^2 \ , 
\label{3-3a}\\
& &\Gamma_{sh^0}=\sinh |\gamma|^2 \ . 
\label{3-3b}
\end{eqnarray}
\end{subequations}
For these two states, we have 
\begin{subequations}\label{3-4}
\begin{eqnarray}
& &{\hat c}\dket{ch^0}=\gamma \dket{sh^0} \ , \qquad
{\hat c}\dket{sh^0}=\gamma\dket{ch^0} \ , 
\label{3-4a}\\
{\hbox{\rm i.e.,}}& & \nonumber\\
& &{\hat \tau}_-\ket{ch^0}=(\gamma^2/2) \ket{ch^0} \ , \qquad
{\hat \tau}_-\ket{sh^0}=(\gamma^2/2)\ket{sh^0} \ . 
\label{3-4b}
\end{eqnarray}
\end{subequations}
The relations (\ref{3-4a}) and (\ref{3-4b}) lead us to 
\begin{subequations}\label{3-5}
\begin{eqnarray}
& &(\tau_+)_{ch^0}=\gamma^{*2}/2 \ , \ \ 
(\tau_-)_{ch^0}=\gamma^{2}/2 \ , \ \ 
(\tau_0)_{ch^0}=|\gamma|^{2}/2\cdot \tanh |\gamma|^2 +1/4\ , \qquad
\label{3-5a}\\
& &(\tau_+)_{sh^0}=\gamma^{*2}/2 \ , \ \ 
(\tau_-)_{sh^0}=\gamma^{2}/2 \ , \ \ 
(\tau_0)_{sh^0}=|\gamma|^{2}/2\cdot \coth |\gamma|^2 +1/4\ . \qquad
\label{3-5b}
\end{eqnarray}
\end{subequations}

The condition (\ref{2-1}) can be expressed for the states $\ket{ch^0}$ 
in the form 
\begin{subequations}\label{3-6}
\begin{eqnarray}
(1/2)(\gamma^*\partial \gamma/\partial c 
-\gamma \partial \gamma^*/\partial c)\tanh |\gamma|^2 
&=& {c}^*/2-i\partial S/\partial c \ , 
\nonumber\\
(1/2)(\gamma^*\partial \gamma/\partial c^* 
-\gamma \partial \gamma^*/\partial c^*)\tanh |\gamma|^2 
&=& -{c}/2-i\partial S/\partial c^* \ . 
\label{3-6a}
\end{eqnarray}
For the state $\ket{sh^0}$, also, we have 
\begin{eqnarray}
(1/2)(\gamma^*\partial \gamma/\partial c 
-\gamma \partial \gamma^*/\partial c)(\coth |\gamma|^2&-&1/|\gamma|^2) 
+\partial [\ln \sqrt{\gamma/\gamma^*}]/\partial c \nonumber\\
&=& {c}^*/2-i\partial S/\partial c \ , 
\nonumber\\
(1/2)(\gamma^*\partial \gamma/\partial c^* 
-\gamma \partial \gamma^*/\partial c^*)(\coth |\gamma|^2&-&1/|\gamma|^2) 
+\partial [\ln \sqrt{\gamma/\gamma^*}]/\partial c^* \nonumber\\
&=& -{c}/2-i\partial S/\partial c^* \ . 
\label{3-6b}
\end{eqnarray}
\end{subequations}
The relations (\ref{3-6a}) gives us the following possible solution :
%and (\ref{3-6b}) are derived 
%for the cases $\ket{ch^0}$ and $\ket{sh^0}$, respectively. 
%For the relation (\ref{3-6a}), we have the following possible form : 
\begin{subequations}\label{3-7}
\begin{eqnarray}
& &S=0 \ , \quad \gamma=\sqrt{\sqrt{2}c}\cdot 
\sqrt[4]{F_{ch^0}(|\gamma|^2)} \ , \quad 
\gamma^*=\sqrt{\sqrt{2}c^*}\cdot \sqrt[4]{F_{ch^0}(|\gamma|^2)} \ , 
\nonumber\\
& &\ \ F_{ch^0}(|\gamma|^2)=|\gamma|^2 \coth |\gamma|^2 \ 
(=1+|\gamma|^4/3-|\gamma|^8/45+\cdots) \ . 
\label{3-7a}
\end{eqnarray}
On the other hand, the relation (\ref{3-6b}) gives us 
\begin{eqnarray}
& &S=i\ln \sqrt{\gamma/\gamma^*} \ , \quad 
\gamma=\sqrt{\sqrt{2}c}\cdot \sqrt[4]{F_{sh^0}(|\gamma|^2)} \ , \quad 
\gamma^*=\sqrt{\sqrt{2}c^*}\cdot \sqrt[4]{F_{sh^0}(|\gamma|^2)} \ , 
\nonumber\\
& &\ \ F_{sh^0}(|\gamma|^2)=|\gamma|^2 /(\coth |\gamma|^2 -|\gamma|^{-2}) 
\nonumber\\
& &\qquad\quad\qquad
 (=3+|\gamma|^4/5-|\gamma|^8/175+\cdots) \ . 
\label{3-7b}
\end{eqnarray}
\end{subequations}
Then, the relations (\ref{3-5a}) and (\ref{3-5b}) leads us to 
\begin{subequations}\label{3-8}
\begin{eqnarray}
& &(\tau_+)_{ch^0}=c^*\sqrt{F_{ch^0}(|\gamma|^2)/2} \ , \qquad
(\tau_-)_{ch^0}=c\sqrt{F_{ch^0}(|\gamma|^2)/2} \ , \nonumber\\
& &(\tau_0)_{ch^0}=|c|^2+1/4 \ , 
\label{3-8a}\\
& &(\tau_+)_{sh^0}=c^*\sqrt{F_{sh^0}(|\gamma|^2)/2} \ , \qquad 
(\tau_-)_{sh^0}=c\sqrt{F_{sh^0}(|\gamma|^2)/2} \ , \nonumber\\ 
& &(\tau_0)_{sh^0}=|c|^2+3/4 \ . \qquad
\label{3-8b}
\end{eqnarray}
\end{subequations}

The relations (\ref{2-3}) and (\ref{2-9}) tell us that for the state 
$\ket{cr^0}$, $(c, c^*)$ plays a role of $({\hat c} , {\hat c}^*)$ 
in the classical mechanics. On the other hand, $(c, c^*)$ for the states 
$\ket{ch^0}$ and $\ket{sh^0}$ plays a role of the boson-pair in the 
classical mechanics. Further, it may be interesting to see in the relations 
(\ref{3-8a}) and (\ref{3-8b}) that the expectation values of the boson 
number ${\hat c}^*{\hat c}$ for the states $\ket{ch^0}$ and $\ket{sh^0}$ 
are expressed as $2c^*c$ and $2c^*c+1$, respectively, which may be 
natural results. 
For the discussion on $(\tau_\pm)_{ch^0}$ and $(\tau_\pm)_{sh^0}$, 
we must give the explicit expressions of $F_{ch^0}(|\gamma|^2)$ and 
$F_{sh^0}(|\gamma|^2)$ in terms of $(c, c^*)$. 
We obtain the relations for $F_{ch^0}(|\gamma|^2)$ and $F_{sh^0}(|\gamma|^2)$ 
with the use of the relations (\ref{3-7a}) and (\ref{3-7b}) : 
\begin{subequations}\label{3-9}
\begin{eqnarray}
& &|\gamma|^4=2|c|^2\cdot F_{ch^0}(|\gamma|^2) \ , 
\quad\hbox{\rm i.e.,}\quad
2|c|^2=|\gamma|^4\cdot F_{ch^0}(|\gamma|^2)^{-1} \ , 
\label{3-9a}\\
& &|\gamma|^4=2|c|^2\cdot F_{sh^0}(|\gamma|^2) \ , 
\quad\hbox{\rm i.e.,}\quad
2|c|^2=|\gamma|^4\cdot F_{sh^0}(|\gamma|^2)^{-1} \ . 
\label{3-9b}
\end{eqnarray}
\end{subequations}
Together with the definitions of $F_{ch^0}(|\gamma|^2)$ and 
$F_{sh^0}(|\gamma|^2)$, we can determine $|\gamma|^2$ as a function of 
$2|c|^2$ for each case. However, 
it is impossible to give the exact analytical expressions. 
Possible approximate forms are shown as follows : 
\begin{subequations}\label{3-10}
\begin{eqnarray}
F_{ch^0}(|\gamma|^2)&=&G_{ch^0}(2|c|^2)+2|c|^2 \ , 
\label{3-10a}\\
F_{sh^0}(|\gamma|^2)&=&G_{sh^0}(2|c|^2)+(2|c|^2+2) \ , \qquad\qquad\quad\qquad
\label{3-10b}
\end{eqnarray}
\end{subequations}
\vspace{-0.8cm}
\begin{subequations}\label{3-11}
\begin{eqnarray}
G_{ch^0}(2|c|^2)
%&\sim& H_{ch^0}(2|c|^2) \nonumber\\
&\sim& 
\exp \left[-(2/3)\cdot 2|c|^2-(2/15)\cdot (2|c|^2)^2\right] 
\ , 
\label{3-11a}\\
G_{sh^0}(2|c|^2)
%&\sim& H_{sh^0}(2|c|^2) \nonumber\\
&\sim& 
\exp \left[-(2/5)\cdot 2|c|^2-(2/175)\cdot (2|c|^2)^2\right] 
\ . 
\label{3-11b}
\end{eqnarray}
\end{subequations}
Derivation of the expressions (\ref{3-10}) and (\ref{3-11}) is shown in 
Appendix A, together with the discussion on the reliability of 
the approximation. 
Through this discussion, it may be expected that the approximate form 
(\ref{3-11}) presents us good agreement with the exact one in the 
quantitative aspect. 

Finally, on the basis of the classical aspect of the $su(1,1)$-algebra, 
we investigate the qualitative feature of the present result. 
For this aim, we define a set $(\tau_{\pm,0}(t))$ in the form 
\begin{equation}\label{3-12}
\tau_+(t)=c^*\sqrt{2t+|c|^2} \ , \quad
\tau_-(t)=c\sqrt{2t+|c|^2} \ , \quad
\tau_0(t)=|c|^2+t \ .
\end{equation}
The relation of the Poisson bracket for the set $(\tau_{\pm,0}(t))$ 
is of the same form as that shown in the relation (\ref{2-5}) :
\setcounter{equation}{11}
\begin{subequations}
\begin{equation}\label{3-12a}
[\tau_+(t) , \tau_-(t) ]_P=(-i)(-2\tau_{0}(t)) \ , \qquad
[\tau_0(t) , \tau_\pm(t) ]_P=(-i)(\pm\tau_{\pm}(t)) \ . 
\end{equation}
\end{subequations}
%The relation (\ref{3-13}) can be regarded as a classical counterpart of 
%the relation (\ref{2-3a}). 
The above tells us that $(\tau_{\pm,0}(t))$ is regarded as a classical 
counterpart of $({\hat \tau}_{\pm,0})$. 
Our present discussion starts in the following inequalities : 
\setcounter{equation}{12}
\begin{subequations}\label{3-13}
\begin{eqnarray}
& &|\gamma|^2 < F_{ch^0}(|\gamma|^2) < \left(\sqrt{1+4|\gamma|^4}+1\right)/2 
\ , 
\label{3-13a}\\
& &\sqrt{1+|\gamma|^4}+1 < F_{sh^0}(|\gamma|^2) < 
\left(\sqrt{9+4|\gamma|^4}+3\right)/2 \ . 
\label{3-13b}
\end{eqnarray}
\end{subequations}
The inequalities (\ref{3-13}) can be easily checked by 
numerical calculation. With the use of the relations (\ref{3-9}), 
the inequalities (\ref{3-13}) can be 
rewritten as 
\begin{subequations}\label{3-14}
\begin{eqnarray}
& &2|c|^2\cdot|\gamma|^2 < |\gamma|^4 < 
2|c|^2\cdot \left(\sqrt{1+4|\gamma|^4}+1\right)/2 \ , 
\label{3-14a}\\
& &2|c|^2\cdot \left(\sqrt{1+|\gamma|^4}+1\right) < |\gamma|^4 < 
2|c|^2\cdot \left(\sqrt{9+4|\gamma|^4}+3\right)/2 \ . 
\label{3-14b}
\end{eqnarray}
\end{subequations}
The above relation gives us 
\begin{subequations}\label{3-15}
\begin{eqnarray}
& &4|c|^2\cdot|c|^2 < |\gamma|^4 < 
4|c|^2\cdot(1/2+|c|^2) \ , 
\label{3-15a}\\
& &4|c|^2\cdot (1+|c|^2) < |\gamma|^4 < 
4|c|^2\cdot (3/2+|c|^2) \ . 
\label{3-15b}
\end{eqnarray}
\end{subequations}
Noting the relations $|(\tau_\pm)_{ch^0}|=|\gamma|^2/2$ and 
$|(\tau_\pm)_{sh^0}|=|\gamma|^2/2$, the inequalities (\ref{3-15}) 
is rewritten as 
\begin{subequations}\label{3-16}
\begin{eqnarray}
& &|\tau_\pm(0)| < |(\tau_\pm)_{ch^0}| < |\tau_\pm(1/4)| \ , 
\label{3-16a}\\
& &|\tau_\pm(1/2)| < |(\tau_\pm)_{sh^0}| < |\tau_\pm(3/4)| \ . 
\label{3-16b}
\end{eqnarray}
\end{subequations}
The quantities $|(\tau_\pm)_{ch^0}|$ and $|(\tau_\pm)_{sh^0}|$ are 
complicated functions of $|c|^2$. However, it may be interesting to see 
that there exist the lower and the upper limits, which are characterized 
by $t=0, 1/4, 1/2$ and $3/4$ contained in the $su(1,1)$-algebra. 
Numerically, we can show the relations 
\begin{subequations}\label{3-17}
\begin{eqnarray}
& &|(\tau_\pm)_{ch^0}| \lsim |\tau_\pm(1/4)| \ , 
\label{3-17a}\\
& &|(\tau_\pm)_{sh^0}| \lsim |\tau_\pm(3/4)| \ , \quad (\hbox{\rm for}\ 
|\gamma|^2 \sim 0) \ , \quad
\label{3-17b}
\end{eqnarray}
\end{subequations}
\vspace{-0.8cm}
\begin{subequations}\label{3-18}
\begin{eqnarray}
& &|\tau_\pm(0)| \lsim |(\tau_\pm)_{ch^0}| \ , 
\label{3-18a}\\
& &|\tau_\pm(1/2)| \lsim |(\tau_\pm)_{sh^0}| \ , \quad (\hbox{\rm for}\ 
|\gamma|^2 \longrightarrow \infty) \ .
\label{3-18b}
\end{eqnarray}
\end{subequations}
The above is our starting form for the deformed boson scheme stressing the 
even-odd boson number difference. However, $(\tau_{\pm,0})_{ch^0}$ and 
$(\tau_{\pm,0})_{sh^0}$ do not satisfy the relation of the Poisson bracket of 
the $su(1,1)$-generators sufficiently. 
However, as is suggested in the relation (\ref{3-4b}), the expectation 
values of the products of $({\hat \tau}_{\pm,0})$ in normal order 
for $({\hat c}, {\hat c}^*)$ are exactly equal to the corresponding products 
of $(\tau_{\pm,0})_{ch^0}$ and $(\tau_{\pm,0})_{sh^0}$.

%\section{The deformation of the states $\ket{ch^0}$ and $\ket{sh^0}$} 
\section{Deformation of the two trial states} 

Main task of this section is to apply the deformed boson scheme 
developed by the present authors\cite{two} 
to the states $\ket{ch^0}$ and 
$\ket{sh^0}$ introduced in \S 3. 
For this aim, we make the deformation of the state $\ket{cr^0}$ 
shown in the relation (\ref{2-6}) in the form 
\begin{equation}\label{4-1}
\ket{tr}=\ket{cr}=\left(\sqrt{\Gamma_{cr}}\right)^{-1}\dket{ex} \ , \qquad
(\bra{cr}cr\rangle =1) 
\end{equation}
\vspace{-0.4cm}
\setcounter{equation}{0}
\begin{subequations}
\begin{equation}\label{4-1a}
\dket{ex}=\exp \left( \gamma {\hat c}^* {\tilde f}({\hat N})\right) \ket{0} \ .
\quad ({\hat N}={\hat c}^*{\hat c})
\end{equation}
\end{subequations}
Here, ${\tilde f}({\hat N})$ is a function of ${\hat N}$ which obeys 
the condition 
\begin{equation}\label{4-2}
{\tilde f}(0)=1 \ , \qquad {\tilde f}(n) > 0 \ . \qquad
(n=1,2,3,\cdots)
\end{equation}
The deformation is characterized by the function ${\tilde f}({\hat N})$ 
and the above is the starting idea of Ref.\citen{two}. 
In the same manner as that in the case of $\dket{ex^0}$, we decompose 
$\dket{ex}$ into two parts : 
\begin{equation}\label{4-3}
\dket{ex}=\dket{ch} +\dket{sh} \ , \qquad\qquad\qquad\qquad\qquad\qquad\qquad
\ \ 
\end{equation}
\vspace{-0.6cm}
\setcounter{equation}{2}
\begin{subequations}
\begin{eqnarray}
\dket{ch}&=& (1/2)\cdot \left(\exp\left(\gamma{\hat c}^*{\tilde f}({\hat N})
\right)+\exp\left(-\gamma{\hat c}^*{\tilde f}({\hat N})\right)\right) \ket{0} 
\nonumber\\
&=&\cosh \left(\gamma{\hat c}^*{\tilde f}({\hat N})\right) \ket{0} \ , 
\label{4-3a}\\
\dket{sh}&=& (1/2)\cdot \left(\exp\left(\gamma{\hat c}^*{\tilde f}({\hat N})
\right)-\exp\left(-\gamma{\hat c}^*{\tilde f}({\hat N})\right)\right) \ket{0} 
\nonumber\\
&=&\sinh \left(\gamma{\hat c}^*{\tilde f}({\hat N})\right) \ket{0} \ . 
\label{4-3b}
\end{eqnarray}
\end{subequations}
In this paper, we will treat the case where ${\tilde f}({\hat N})$ 
does not depend on any other operator and parameter. 
It may be self-evident that $\dket{ch}$ and $\dket{sh}$ consist of 
the states with the even- and the odd-boson numbers, respectively. 
The states $\dket{ch}$ and $\dket{sh}$ can be expressed explicitly 
in the form 
\begin{subequations}\label{4-4}
\begin{eqnarray}
& &\dket{ch}=\ket{0}+\sum_{n=1}^{\infty}
\frac{\gamma^{2n}}{\sqrt{(2n)!}}{\tilde f}(0)\cdots {\tilde f}(2n-1)
\ket{2n} \ , 
\label{4-4a}\\
& &\dket{sh}=\sum_{n=0}^{\infty}
\frac{\gamma^{2n+1}}{\sqrt{(2n+1)!}}{\tilde f}(0)\cdots {\tilde f}(2n)
\ket{2n+1} \ . 
\label{4-4b}
\end{eqnarray}
\end{subequations}
Here, $\ket{k}$ ($k=2n, 2n+1$) denotes 
\begin{equation}\label{4-5}
\ket{k}=(1/\sqrt{k!})\cdot ({\hat c}^*)^k \ket{0} \ . \qquad 
(\bra{k}k\rangle =1) 
\end{equation}
Including the normalization constants, we define 
\begin{subequations}\label{4-6}
\begin{eqnarray}
& &\ket{tr}=\ket{ch}=\left(\sqrt{\Gamma_{ch}}\right)^{-1} \dket{ch} \ , 
\qquad (\bra{ch} ch \rangle =1) 
\label{4-6a}\\
& &\ket{tr}=\ket{sh}=\left(\sqrt{\Gamma_{sh}}\right)^{-1} \dket{sh} \ , 
\qquad (\bra{sh} sh \rangle =1) 
\label{4-6b}
\end{eqnarray}
\end{subequations}
\vspace{-0.6cm}
\begin{subequations}\label{4-7}
\begin{eqnarray}
& &\Gamma_{ch}=1+\sum_{n=1}^{\infty}
\frac{(|\gamma|^2)^{2n}}{(2n)!}({\tilde f}(0)\cdots {\tilde f}(2n-1))^2 \ ,
\label{4-7a}\\
& &\Gamma_{sh}=\sum_{n=0}^{\infty}
\frac{(|\gamma|^2)^{2n+1}}{(2n+1)!}({\tilde f}(0)\cdots {\tilde f}(2n))^2 \ .
\label{4-7b}
\end{eqnarray}
\end{subequations}
For these two states, we have the relation 
\begin{subequations}\label{4-8}
\begin{equation}\label{4-8a}
{\tilde f}({\hat N})^{-1}{\hat c}\dket{ch}
=\gamma \dket{sh} \ , \quad
{\tilde f}({\hat N})^{-1}{\hat c}\dket{sh}=\gamma\dket{ch} \ , 
\end{equation}
that is, 
\begin{equation}\label{4-8b}
\left({\tilde f}({\hat N})^{-1}{\hat c}\right)^2\ket{ch}
=\gamma^2\ket{ch} \ , \quad
\left({\tilde f}({\hat N})^{-1}{\hat c}\right)^2\ket{sh}
=\gamma^2\ket{sh} \ .
\end{equation}
The relation (\ref{4-8b}) can be rewritten in the form 
\begin{eqnarray}\label{4-8c}
& &\left[ {\tilde f}(2{\hat \tau}_0-1/2){\tilde f}(2{\hat \tau}_0+1/2)
\right]^{-1}{\hat \tau}_- \ket{ch}=(\gamma^2/2)\ket{ch} \ , \nonumber\\
& &\left[ {\tilde f}(2{\hat \tau}_0-1/2){\tilde f}(2{\hat \tau}_0+1/2)
\right]^{-1}{\hat \tau}_- \ket{sh}=(\gamma^2/2)\ket{sh} \ . 
\end{eqnarray}
\end{subequations}
Here, we used the formula 
\begin{equation}\label{4-9}
[\ {\tilde f}({\hat N})^{-1}{\hat c} \ , \ {\hat c}^*{\tilde f}({\hat N})\ ]
=1\ . 
\end{equation}
The relation (\ref{4-2}) supports the existence of 
${\tilde f}({\hat N})^{-1}$. 
With the use of the relations (\ref{4-8}), we get 
\begin{subequations}\label{4-10}
\begin{eqnarray}
& &(\tau_+)_{ch}=\gamma^{*2}/2\cdot 
%\bra{ch}
\left({\tilde f}({N}){\tilde f}({N}+1) \right)_{ch}
%\ket{ch} 
\ , 
\nonumber\\
& &(\tau_-)_{ch}=\gamma^{2}/2\cdot 
%\bra{ch}
\left({\tilde f}({N}){\tilde f}({N}+1)\right)_{ch} 
%\ket{ch} 
\ , 
\nonumber\\
& &(\tau_0)_{ch}=|\gamma|^{2}/2\cdot \left(\Gamma'/\Gamma\right)_{ch}
+1/4 \ , 
\label{4-10a}\\
& &(\tau_+)_{sh}=\gamma^{*2}/2\cdot \left({\tilde f}({N})
{\tilde f}({N}+1)\right)_{sh} \ , 
\nonumber\\
& &(\tau_-)_{sh}=\gamma^{2}/2\cdot \left({\tilde f}({N})
{\tilde f}({N}+1)\right)_{sh} \ , 
\nonumber\\
& &(\tau_0)_{sh}=|\gamma|^{2}/2\cdot \left(\Gamma'/\Gamma\right)_{sh} 
+1/4 \ . 
\label{4-10b}
\end{eqnarray}
\end{subequations}
Here, $({\tilde f}(N){\tilde f}(N+1))_{ch}$, $(\Gamma'/\Gamma)_{ch}$, 
$({\tilde f}(N){\tilde f}(N+1))_{sh}$ and $(\Gamma'/\Gamma)_{sh}$ are 
defined as 
\begin{subequations}\label{4-11}
\begin{eqnarray}
& &\left({\tilde f}(N){\tilde f}(N+1)\right)_{ch}
=\bra{ch} {\tilde f}({\hat N}){\tilde f}({\hat N}+1)\ket{ch} \ , 
\nonumber\\
& &\left(\Gamma'/\Gamma\right)_{ch}=\left(d\Gamma_{ch}/d|\gamma|^2\right)
\cdot \left(\Gamma_{ch}\right)^{-1} \ , 
\label{4-11a}\\
& &\left({\tilde f}(N){\tilde f}(N+1)\right)_{sh}
=\bra{sh} {\tilde f}({\hat N}){\tilde f}({\hat N}+1)\ket{sh} \ , 
\nonumber\\
& &\left(\Gamma'/\Gamma\right)_{sh}=\left(d\Gamma_{sh}/d|\gamma|^2\right)
\cdot \left(\Gamma_{sh}\right)^{-1} \ . 
\label{4-11b}
\end{eqnarray}
\end{subequations}

The condition (\ref{2-1}) in the present case is expressed in 
the following form : 
\begin{subequations}\label{4-12}
\begin{eqnarray}
(1/2)\cdot(\gamma^* \partial \gamma/\partial c -
\gamma \partial \gamma^*/\partial c)\cdot 
\left(\Gamma'/\Gamma\right)_{ch}
&=& {c}^*/2 -i \partial S/\partial c \ , 
\nonumber\\
(1/2)\cdot(\gamma^* \partial \gamma/\partial c^* -
\gamma \partial \gamma^*/\partial c^*)\cdot 
\left(\Gamma'/\Gamma\right)_{ch}
&=& -{c}/2 -i \partial S/\partial c^* \ , 
\label{4-12a}\\
(1/2)\cdot(\gamma^* \partial \gamma/\partial c -
\gamma \partial \gamma^*/\partial c)\cdot 
\biggl(\left(\Gamma'/\Gamma\right)_{sh}&-&1/|\gamma|^2\biggl)
\nonumber\\
+\partial [\ln \sqrt{\gamma/\gamma^*}]/\partial c
&=& {c}^*/2 -i \partial S/\partial c \ , 
\nonumber\\
(1/2)\cdot(\gamma^* \partial \gamma/\partial c^* -
\gamma \partial \gamma^*/\partial c^*)\cdot 
\biggl(\left(\Gamma'/\Gamma\right)_{sh}&-&1/|\gamma|^2\biggl)
\nonumber\\
+\partial [\ln \sqrt{\gamma/\gamma^*}]/\partial c^*
&=& -{c}/2 -i \partial S/\partial c^* \ . \qquad\quad
\label{4-12b}
\end{eqnarray}
\end{subequations}
For the relation (\ref{4-12}), we have the following possible form : 
\begin{subequations}\label{4-13}
\begin{eqnarray}
& &S=0 \ , \quad 
\gamma=\sqrt{\sqrt{2}c}\cdot \sqrt[4]{F_{ch}(|\gamma|^2)} \ , 
\quad
\gamma^*=\sqrt{\sqrt{2}c^*}\cdot \sqrt[4]{F_{ch}(|\gamma|^2)} \ , 
\nonumber\\
& &F_{ch}(|\gamma|^2)=|\gamma|^2\cdot \left(\Gamma'/\Gamma\right)_{ch}^{-1}
\ , 
\label{4-13a}\\
& &S=i\ln \sqrt{\gamma/\gamma^*} \ , \quad 
\gamma=\sqrt{\sqrt{2}c}\cdot \sqrt[4]{F_{sh}(|\gamma|^2)} \ , 
\quad
\gamma^*=\sqrt{\sqrt{2}c^*}\cdot \sqrt[4]{F_{sh}(|\gamma|^2)} \ , 
\nonumber\\
& &F_{sh}(|\gamma|^2)=|\gamma|^2/\left[ 
\left(\Gamma'/\Gamma\right)_{sh}-1/|\gamma|^2 \right]^{-1} \ . 
\label{4-13b}
\end{eqnarray}
\end{subequations}
The relations (\ref{4-13}) give us the following 
relations for expressing $|\gamma|^2$ in terms of $2|c|^2$ : 
\begin{subequations}\label{4-14}
\begin{eqnarray}
& &|\gamma|^4=2|c|^2\cdot F_{ch}(|\gamma|^2) \ , 
\label{4-14a}\\
& &|\gamma|^4=2|c|^2\cdot F_{sh}(|\gamma|^2) \ . 
\label{4-14b}
\end{eqnarray}
\end{subequations}
By solving the relations (\ref{4-14}), we can 
express $|\gamma|^2$ in terms of $2|c|^2$ and, then, substituting 
the results into $F_{ch}(|\gamma|^2)$ and $F_{sh}(|\gamma|^2)$, 
we can express $F_{ch}(|\gamma|^2)$ and $F_{sh}(|\gamma|^2)$ in 
terms of $2|c|^2$. Thus, we have $(\gamma, \gamma^*)$ which is expressed 
as a function of $(c, c^*)$. 
Further, if $({\tilde f}({N}){\tilde f}({N}+1))_{ch}$ 
and $({\tilde f}({N}){\tilde f}({N}+1))_{sh}$ are 
calculated, $(\tau_\pm)_{ch}$ and $(\tau_\pm)_{sh}$ shown in the 
relations (\ref{4-10a}) and (\ref{4-10b}) can be expressed in terms of 
$(c, c^*)$. 
Concerning $(\tau_0)_{ch}$ and $(\tau_0)_{sh}$, in any case, we have 
\begin{subequations}\label{4-15}
\begin{eqnarray}
& &(\tau_0)_{ch}=|c|^2 +1/4 \ , 
\label{4-15a}\\
& &(\tau_0)_{sh}=|c|^2 +3/4 \ . 
\label{4-15b}
\end{eqnarray}
\end{subequations}
With the use of the relations (\ref{4-10}) and (\ref{4-13}), we can derive 
the form (\ref{4-15}). 
It is impossible in the general framework to mention definitely if the 
relation of the Poisson bracket (\ref{2-5}) is satisfied or not. 
Further, as is clear from the relation (\ref{4-8c}), generally, 
the expectation 
values of the products of $({\hat \tau}_{\pm,0})$ in normal order for 
$({\hat c} , {\hat c}^*)$ are not equal to the corresponding products of 
$(\tau_{\pm,0})_{ch}$ and $(\tau_{\pm,0})_{sh}$.

Finally, we contact with the connection of the form presented in this 
section with the original one given in \S 3. 
It may be self-evident that, if ${\tilde f}({\hat N})=1$, the form 
in this section is completely reduced to the original one in \S 3. 
We investigate this reduction from rather wider viewpoint. 
First, we treat the following case : 
\begin{equation}\label{4-16}
{\tilde f}(0){\tilde f}(1)\cdots {\tilde f}(2n-2){\tilde f}(2n-1)=1 \ . 
\end{equation}
Then, $\dket{ch}$ and $\dket{sh}$ are expressed as 
\begin{subequations}\label{4-17}
\begin{eqnarray}
& &\dket{ch}=\dket{ch^0} \ , 
\label{4-17a}\\
& &\dket{sh}=\sum_{n=0}^{\infty} \gamma^{2n+1}
({\tilde f}(2n)/\sqrt{(2n+1)!}) \ket{2n+1} \ , 
\label{4-17b}
\end{eqnarray}
\end{subequations}
\vspace{-0.3cm}
\begin{subequations}\label{4-18}
\begin{eqnarray}
& &\Gamma_{ch}=\Gamma_{ch^0} \ , 
\label{4-18a}\\
& &\Gamma_{sh}=\sum_{n=0}^{\infty} (|\gamma|^2)^{2n+1}
({\tilde f}(2n)^2/(2n+1)!) \ . 
\label{4-18b}
\end{eqnarray}
\end{subequations}
The relations (\ref{4-17}) and (\ref{4-18}) tell us that, depending 
on the choice of ${\tilde f}(2n)$, the state $\ket{ch^0}$, the 
partner of which is not reduced to the state $\ket{sh^0}$, 
should be regarded as a 
possible form of the $q$-deformation. 
An example which satisfies the relation (\ref{4-16}) is given as 
\begin{equation}\label{4-19}
{\tilde f}(2n){\tilde f}(2n+1)=1 \ . \qquad
(n=0,1,2,\cdots)
\end{equation}
In order to understand the above statement, we investigate the case 
\begin{equation}\label{4-20}
{\tilde f}(2n)=\sqrt{1-\alpha+\alpha(2n+1)} \ . 
\quad (\alpha \ge 0) 
\end{equation}
Of course, we define ${\tilde f}(2n+1)={\tilde f}(2n)^{-1}$. 
In this case, we have 
\begin{equation}\label{4-21}
\Gamma_{sh}=(1-\alpha)\Gamma_{sh^0}+\alpha \cdot |\gamma|^2 \Gamma_{ch^0} \ . 
\end{equation}
Clearly, if $\alpha=0$, $\Gamma_{sh}=\Gamma_{sh^0}$ and if $\alpha=1$, 
$\Gamma_{sh}=|\gamma|^2\Gamma_{ch^0}$. 
Next, we treat the case 
\begin{equation}\label{4-22}
{\tilde f}(0){\tilde f}(1)\cdots {\tilde f}(2n-1){\tilde f}(2n)=1 \ . 
\end{equation}
In this case, $\dket{ch}$ and $\dket{sh}$ are obtained in the form 
\begin{subequations}\label{4-23}
\begin{eqnarray}
& &\dket{ch}=\sum_{n=0}^{\infty}\gamma^{2n}
({\tilde f}(2n)^{-1}/\sqrt{(2n)!})\ket{2n} \ , 
\label{4-23a}\\
& &\dket{sh}=\dket{sh^0} \ , 
\label{4-23b}
\end{eqnarray}
\end{subequations}
\vspace{-0.6cm}
\begin{subequations}\label{4-24}
\begin{eqnarray}
& &\Gamma_{ch}=\sum_{n=0}^{\infty}(|\gamma|^2)^{2n}
({\tilde f}(2n)^{-2}/(2n)!) \ , 
\label{4-24a}\\
& &\Gamma_{sh}=\Gamma_{sh^0} \ . 
\label{4-24b}
\end{eqnarray}
\end{subequations}
This case is also in the same situation as that in the case (\ref{4-16}). 
A possible choice of the relation 
\begin{equation}\label{4-25}
{\tilde f}(2n-1){\tilde f}(2n)=1 \ . \qquad
(n=1,2,3,\cdots)
\end{equation}
An illustrative example is shown in the case 
\begin{equation}\label{4-26}
{\tilde f}(2n)^{-1}=\sqrt{1-\beta+\beta(2n+1)^{-1}} \ . 
\quad (\beta \le 1) 
\end{equation}
In this case, we have 
\begin{equation}\label{4-27}
\Gamma_{ch}=(1-\beta)\Gamma_{ch^0}+\beta \cdot |\gamma|^{-2} \Gamma_{sh^0} \ . 
\end{equation}
Clearly, if $\beta=0$, $\Gamma_{ch}=\Gamma_{ch^0}$ and if $\beta=1$, 
$\Gamma_{ch}=|\gamma|^{-2}\Gamma_{sh^0}$. 
In the next section, we discuss the meaning of 
$\Gamma_{sh}=|\gamma|^2\Gamma_{ch^0}$ and 
$\Gamma_{ch}=|\gamma|^{-2}\Gamma_{sh^0}$.

\section{Various examples}

In this section, we show various examples of the general framework 
developed in \S 4. 
First, we rewrite $\dket{ch}$ and $\dket{sh}$ shown in 
the relations (\ref{4-4a}) and (\ref{4-4b}) in 
the following forms : 
\begin{subequations}\label{5-1}
\begin{eqnarray}
& &\dket{ch}=\gamma^{-1}\left(\sqrt{{\hat N}+1}\right)^{-1}{\hat c}
\sum_{n=0}^{\infty}\frac{\gamma^{2n+1}}{\sqrt{(2n+1)!}}{\tilde f}(0)\cdots
{\tilde f}(2n)\frac{\sqrt{2n+1}}{{\tilde f}(2n)}\ket{2n+1} \ , \nonumber\\
& &
\label{5-1a}\\
& &\dket{sh}=\gamma{\hat c}^*\left(\sqrt{{\hat N}+1}\right)^{-1}
\left(\ket{0}+
\sum_{n=1}^{\infty}\frac{\gamma^{2n}}{\sqrt{(2n)!}}{\tilde f}(0)\cdots
{\tilde f}(2n-1)\frac{{\tilde f}(2n)}{\sqrt{2n+1}}\ket{2n}\right) \ , 
\nonumber\\
& &
\label{5-1b}
\end{eqnarray}
\end{subequations}
\vspace{-0.4cm}
\begin{subequations}\label{5-2}
\begin{eqnarray}
& &\Gamma_{ch}=|\gamma|^{-2}
\sum_{n=0}^{\infty}\frac{(|\gamma|^2)^{2n+1}}{(2n+1)!}\left({\tilde f}(0)\cdots
{\tilde f}(2n)\right)^2\frac{{2n+1}}{{\tilde f}(2n)^2} \ , 
\label{5-2a}\\
& &\Gamma_{sh}=|\gamma|^2
\left(1+
\sum_{n=1}^{\infty}\frac{(|\gamma|^2)^{2n}}{{(2n)!}}\left({\tilde f}(0)\cdots
{\tilde f}(2n-1)\right)^2\frac{{\tilde f}(2n)^2}{{2n+1}}\right) \ . 
\label{5-2b}
\end{eqnarray}
\end{subequations}
To the above expressions, we require the following condition ; 
\begin{equation}\label{5-3}
{\tilde f}(2n)=\sqrt{2n+1} \ . 
\end{equation}
Then, we have 
\begin{subequations}\label{5-4}
\begin{eqnarray}
& &\dket{ch}=\gamma^{-1}\left(\sqrt{{\hat N}+1}\right)^{-1} {\hat c} 
\dket{sh} \ , 
\label{5-4a}\\
& &\dket{sh}=\gamma{\hat c}^*\left(\sqrt{{\hat N}+1}\right)^{-1} 
\dket{ch} \ , 
\label{5-4b}
\end{eqnarray}
\end{subequations}
\vspace{-0.4cm}
\begin{subequations}\label{5-5}
\begin{eqnarray}
& &\Gamma_{ch}=|\gamma|^{-2}\Gamma_{sh} \ , 
\label{5-5a}\\
& &\Gamma_{sh}=|\gamma|^2 \Gamma_{ch} \ . 
\label{5-5b}
\end{eqnarray}
\end{subequations}
Therefore, the following forms are derived : 
\begin{subequations}\label{5-6}
\begin{eqnarray}
& &\ket{ch}=
(\gamma/|\gamma|)^{-1}\left(\sqrt{{\hat N}+1}\right)^{-1}{\hat c}\dket{sh} 
\ , 
\label{5-6a}\\
& &\ket{sh}
=(\gamma/|\gamma|){\hat c}^*\left(\sqrt{{\hat N}+1}\right)^{-1}\dket{ch} \ . 
\label{5-6b}
\end{eqnarray}
\end{subequations}
Here, $(\sqrt{{\hat N}+1})^{-1}{\hat c}$ and 
${\hat c}^*(\sqrt{{\hat N}+1})^{-1}$ obey 
\begin{eqnarray}\label{5-7}
& &\left(\sqrt{{\hat N}+1}\right)^{-1}{\hat c}\cdot 
{\hat c}^*\left(\sqrt{{\hat N}+1}\right)^{-1}
=1 \ , \nonumber\\
& &{\hat c}^*\left(\sqrt{{\hat N}+1}\right)^{-1}\cdot 
\left(\sqrt{{\hat N}+1}\right)^{-1}{\hat c}
=1-\ket{0}\bra{0} \ .
\end{eqnarray}
The relation (\ref{5-7}) tells us that $(\sqrt{{\hat N}+1})^{-1}{\hat c}$ and 
${\hat c}^*(\sqrt{{\hat N}+1})^{-1}$ play a role of the phase operators and 
the form (\ref{5-6}) suggests us that, under the condition 
(\ref{5-3}), the unpaired boson does not give any influence on the 
boson-pairs. Further, the meaning of 
$\Gamma_{sh}=|\gamma|^2\Gamma_{ch^0}$ and 
$\Gamma_{ch}=|\gamma|^{-2}\Gamma_{sh^0}$ appearing in \S 4 
may be understandable. 

Next, we investigate the case in which analytical and concrete 
expressions are given. This case starts in the following condition :
\begin{equation}\label{5-8}
{\tilde f}(0)\cdots {\tilde f}(2n-1)=\sqrt{(2n)!} \ . 
\end{equation}
Then, $\dket{ch}$ and $\dket{sh}$ are of the forms 
\begin{subequations}\label{5-9}
\begin{eqnarray}
& &\dket{ch}=\sum_{n=0}^{\infty}\gamma^{2n}\ket{2n} \ , 
\label{5-9a}\\
& &\dket{sh}=\sum_{n=0}^{\infty}\gamma^{2n+1}\left({\tilde f}(2n)
/\sqrt{2n+1}\right)\ket{2n+1} \ , 
\label{5-9b}
\end{eqnarray}
\end{subequations}
\vspace{-0.4cm}
\begin{subequations}\label{5-10}
\begin{eqnarray}
& &\Gamma_{ch}=\sum_{n=0}^{\infty}(|\gamma|^4)^n \ , 
\label{5-10a}\\
& &\Gamma_{sh}=|\gamma|^2\sum_{n=0}^{\infty}(|\gamma|^4)^n 
{\tilde f}(2n)^2/(2n+1) \ . 
\label{5-10b}
\end{eqnarray}
\end{subequations}
In the above case, if ${\tilde f}(2n)=\sqrt{2n+1}$, 
the form is reduced to the case discussed at the beginning part of 
this section. We discuss the following case : 
\begin{equation}\label{5-11}
{\tilde f}(2n)=\sqrt{(n+1)(2n+1)} \ , \qquad
{\tilde f}(2n-1)=\sqrt{2} \ .
\end{equation}
The case (\ref{5-11}) satisfies the condition (\ref{5-8}) and 
${\tilde f}(0)=1$. 
Then, $\Gamma_{ch}$ and $\Gamma_{sh}$ are given as 
\begin{subequations}\label{5-12}
\begin{eqnarray}
& &\Gamma_{ch}=(1-|\gamma|^4)^{-1} \ , 
\label{5-12a}\\
& &\Gamma_{sh}=|\gamma|^2\cdot (1-|\gamma|^4)^{-2} \ .
\label{5-12b}
\end{eqnarray}
\end{subequations}
The states $\ket{ch}$ and $\ket{sh}$ satisfy 
\begin{subequations}\label{5-13}
\begin{eqnarray}
& &\left(\sqrt{(2{\hat \tau}_0+1/2)(2{\hat \tau}_0+3/2)}\right)^{-1}
{\hat \tau}_- \ket{ch}=(\gamma^2/2)\ket{ch} \ , 
\label{5-13a}\\
& &\left(\sqrt{(2{\hat \tau}_0+5/2)(2{\hat \tau}_0+7/2)}\right)^{-1}
{\hat \tau}_- \ket{sh}=(\gamma^2/2)\ket{sh} \ . 
\label{5-13b}
\end{eqnarray}
\end{subequations}
With the use of the form (\ref{5-12}) with the relation (\ref{4-13}), 
we have 
\begin{subequations}\label{5-14}
\begin{eqnarray}
& &F_{ch}(|\gamma|^2)=(1/2)(1-|\gamma|^4) \ , 
\label{5-14a}\\
& &F_{sh}(|\gamma|^2)=(1/4)(1-|\gamma|^4) \ . 
\label{5-14b}
\end{eqnarray}
\end{subequations}
For the forms (\ref{5-14a}) and (\ref{5-14b}), respectively, 
$|\gamma|^4$ can be expressed as 
\begin{subequations}\label{5-15}
\begin{eqnarray}
& &|\gamma|^4=|c|^2\cdot (1+|c|^2)^{-1} \ , \quad \hbox{\rm i.e.,}\quad
F_{ch}(|\gamma|^2)=(1/2)(1+|c|^2)^{-1} \ , 
\label{5-15a}\\
& &|\gamma|^4=|c|^2\cdot (2+|c|^2)^{-1} \ , \quad \hbox{\rm i.e.,}\quad
F_{sh}(|\gamma|^2)=(1/2)(2+|c|^2)^{-1} \ . 
\label{5-15b}
\end{eqnarray}
\end{subequations}
Therefore, the relations (\ref{4-13a}) and (\ref{4-13b}) give us 
\begin{subequations}\label{5-16}
\begin{eqnarray}
& &\gamma=\sqrt{c}\cdot \left(\sqrt[4]{1+|c|^2}\right)^{-1} \ , 
\quad
\gamma^*=\sqrt{c^*}\cdot \left(\sqrt[4]{1+|c|^2}\right)^{-1} \ , 
\label{5-16a}\\
& &\gamma=\sqrt{c}\cdot \left(\sqrt[4]{2+|c|^2}\right)^{-1} \ , 
\quad
\gamma^*=\sqrt{c^*}\cdot \left(\sqrt[4]{2+|c|^2}\right)^{-1} \ . 
\label{5-16b}
\end{eqnarray}
\end{subequations}

The second case starts in the condition 
\begin{equation}\label{5-17}
{\tilde f}(0)\cdots {\tilde f}(2n)=\sqrt{(2n+1)!} \ .
\end{equation}
Then, $\dket{ch}$ and $\dket{sh}$ are given as 
\begin{subequations}\label{5-18}
\begin{eqnarray}
& &\dket{ch}=\sum_{n=0}^{\infty}\gamma^{2n}\left((2n+1)/{\tilde f}(2n)^2
\right) \ket{2n} \ , 
\label{5-18a}\\
& &\dket{sh}=\sum_{n=0}^{\infty}\gamma^{2n+1}\ket{2n+1} \ , 
\label{5-18b}
\end{eqnarray}
\end{subequations}
\vspace{-0.4cm}
\begin{subequations}\label{5-19}
\begin{eqnarray}
& &\Gamma_{ch}=\sum_{n=0}^{\infty}(|\gamma|^4)^n (2n+1)/{\tilde f}(2n)^2 \ , 
\label{5-19a}\\
& &\Gamma_{sh}=\sum_{n=0}^{\infty}(|\gamma|^2)^{2n+1} \ . 
\label{5-19b}
\end{eqnarray}
\end{subequations}
In a manner similar to the form (\ref{5-11}), we discuss the case 
\begin{equation}\label{5-20}
{\tilde f}(2n)=\sqrt{(n+1)^{-1}(2n+1)} \ , \quad
{\tilde f}(2n-1)=\sqrt{2n(n+1)} \ . 
\end{equation}
Of course, the form (\ref{5-20}) satisfies the condition (\ref{5-17}) 
and ${\tilde f}(0)=1$. In a way similar to the previous case, 
we have the following results : 
\begin{subequations}\label{5-21}
\begin{eqnarray}
& &\Gamma_{ch}=(1-|\gamma|^4)^{-2} \ , 
\label{5-21a}\\
& &\Gamma_{sh}=|\gamma|^2\cdot (1-|\gamma|^4)^{-1} \ , \qquad
\label{5-21b}
\end{eqnarray}
\end{subequations}
\vspace{-0.8cm}
\begin{subequations}\label{5-22}
\begin{eqnarray}
& &F_{ch}(|\gamma|^2)=(1/4)\cdot(1-|\gamma|^4) \ , 
\label{5-22a}\\
& &F_{sh}(|\gamma|^2)=(1/2)\cdot(1-|\gamma|^4) \ . 
\label{5-22b}
\end{eqnarray}
\end{subequations}
In this case, the states $\ket{ch}$ and $\ket{sh}$ satisfy the same relation 
as that shown in the relation (\ref{5-13}). 
With the use of the above results, for the states $\ket{ch}$ and 
$\ket{sh}$, we have, respectively, 
\begin{subequations}\label{5-23}
\begin{eqnarray}
& &|\gamma|^4=|c|^2\cdot(2+|c|^2)^{-1} \ , 
\label{5-23a}\\
& &|\gamma|^4=|c|^2\cdot(1+|c|^2)^{-1} \ , 
\label{5-23b}\qquad\qquad
\end{eqnarray}
\end{subequations}
\vspace{-0.6cm}
\begin{subequations}\label{5-24}
\begin{eqnarray}
& &\gamma=\sqrt{c}\cdot \left(\sqrt[4]{2+|c|^2}\right)^{-1} \ , 
\quad
\gamma^*=\sqrt{c^*}\cdot \left(\sqrt[4]{2+|c|^2}\right)^{-1} \ , 
\label{5-24a}\\
& &\gamma=\sqrt{c}\cdot \left(\sqrt[4]{1+|c|^2}\right)^{-1} \ , 
\quad
\gamma^*=\sqrt{c^*}\cdot \left(\sqrt[4]{1+|c|^2}\right)^{-1} \ . 
\label{5-24b}
\end{eqnarray}
\end{subequations}
Thus, $(\gamma , \gamma^*)$ can be expressed in terms of $(c , c^*)$ 
analytically.

Our final task for the deformations (\ref{5-11}) and (\ref{5-20}) 
is to give $|(\tau_\pm)_{ch}|$ and $|(\tau_{\pm})_{sh}|$. 
For this aim, first, we must calculate 
$({\tilde f}({N}){\tilde f}({N}+1))_{ch}$ and 
$({\tilde f}({N}){\tilde f}({N}+1))_{sh}$, 
which have been given in the relations (\ref{4-10a}) and (\ref{4-10b}). 
For the deformation (\ref{5-11}), we have the following expression : 
\begin{subequations}\label{5-25}
\begin{eqnarray}
& &({\tilde f}({N}){\tilde f}({N}+1))_{ch}
=(1-|\gamma|^4)\left(\sum_{n=0}^{\infty}
(|\gamma|^4)^n\sqrt{(2n+1)(2n+2)}\right) \ , 
\label{5-25a}\\
& &({\tilde f}({N}){\tilde f}({N}+1))_{sh}
=(1-|\gamma|^4)^2 \frac{d}{d|\gamma|^4}\left(\sum_{n=0}^{\infty}
(|\gamma|^4)^{n+1}\sqrt{(2n+3)(2n+4)}\right) \ . \nonumber\\
& &
\label{5-25b}
\end{eqnarray}
\end{subequations}
For the deformation (\ref{5-20}), we have 
\begin{subequations}\label{5-26}
\begin{eqnarray}
& &({\tilde f}({N}){\tilde f}({N}+1))_{ch}
=(1-|\gamma|^4)^2 \frac{d}{d|\gamma|^4}\left(\sum_{n=0}^{\infty}
(|\gamma|^4)^{n+1}\sqrt{(2n+1)(2n+4)}\right) \ , \nonumber\\
& &
\label{5-26a}\\
& &({\tilde f}({N}){\tilde f}({N}+1))_{sh}
=(1-|\gamma|^4)\left(\sum_{n=0}^{\infty}
(|\gamma|^4)^{n}\sqrt{(2n+2)(2n+3)}\right) \ . 
\label{5-26b}
\end{eqnarray}
\end{subequations}
It may be impossible to express the above power series expansions in 
compact forms. Then, we note the following inequalities for $x$ 
($x\ge 0$) : 
\begin{eqnarray}\label{5-27}
%& &\sqrt{2}+2x < \sqrt{(2x+1)(2x+2)} < 3/2+2x \ , \nonumber\\
%& &2\sqrt{3}+2x < \sqrt{(2x+3)(2x+4)} < 7/2+2x \ , \nonumber\\
%& &2+2x < \sqrt{(2x+1)(2x+4)} < 5/2+2x \ , \nonumber\\
%& &\sqrt{6}+2x < \sqrt{(2x+3)(2x+2)} < 5/2+2x \ . 
& &2x+\sqrt{\lambda\mu} < \sqrt{(2x+\lambda)(2x+\mu)} < 
2x+(\lambda+\mu)/2 \ . \qquad (\lambda, \mu >0) 
\end{eqnarray}
Further, we have 
\begin{eqnarray}\label{5-28}
& &(1-|\gamma|^4)\sum_{n=0}^{\infty}(|\gamma|^4)^n = 1 \ , \nonumber\\
& &(1-|\gamma|^4)\sum_{n=0}^{\infty}(|\gamma|^4)^n n 
= |\gamma|^4(1-|\gamma|^4)^{-1} \ , \nonumber\\
& &(1-|\gamma|^4)^2 \frac{d}{d|\gamma|^4} 
\sum_{n=0}^{\infty}(|\gamma|^4)^{n+1} = 1 \ , \nonumber\\
& &(1-|\gamma|^4)^2 \frac{d}{d|\gamma|^4} 
\sum_{n=0}^{\infty}(|\gamma|^4)^{n+1}n = 
2|\gamma|^4(1-|\gamma|^4)^{-1} \ . 
\end{eqnarray}
By putting $x=n$, $\lambda=1, 2$ and $\mu=3, 4$ in the inequality 
(\ref{5-27}) and using the relation (\ref{5-28}), we can derive 
the following inequality 
for the quantity (\ref{5-25}) : 
\begin{subequations}\label{5-29}
\begin{eqnarray}
& &\sqrt{2}+2|\gamma|^4(1-|\gamma|^4)^{-1} 
< ({\tilde f}({N}){\tilde f}({N}+1))_{ch}
< 3/2+2|\gamma|^4(1-|\gamma|^4)^{-1} \ , \nonumber\\
& &
\label{5-29a}\\
& &2\sqrt{3}+4|\gamma|^4(1-|\gamma|^4)^{-1}
< ({\tilde f}({N}){\tilde f}({N}+1))_{sh}
< 7/2+4|\gamma|^4(1-|\gamma|^4)^{-1} \ , \nonumber\\
& &
\label{5-29b}
\end{eqnarray}
\end{subequations}
In the same manner as the above, we have the following inequality for 
the quantity (\ref{5-26}) : 
\begin{subequations}\label{5-30}
\begin{eqnarray}
& &2+4|\gamma|^4(1-|\gamma|^4)^{-1} 
< ({\tilde f}({N}){\tilde f}({N}+1))_{ch}
< 5/2+4|\gamma|^4(1-|\gamma|^4)^{-1} \ , \nonumber\\
& &\label{5-30a}\\
& &\sqrt{6}+2|\gamma|^4(1-|\gamma|^4)^{-1} 
< ({\tilde f}({N}){\tilde f}({N}+1))_{sh}
< 5/2+2|\gamma|^4(1-|\gamma|^4)^{-1} \ . \nonumber\\
& &\label{5-30b}
\end{eqnarray}
\end{subequations}
The use of the relations (\ref{5-15}) and (\ref{5-23}) for the 
inequalities (\ref{5-29}) and (\ref{5-30}), respectively, gives us 
\begin{subequations}\label{5-31}
\begin{eqnarray}
& &\sqrt{2}+2|c|^2
< ({\tilde f}({N}){\tilde f}({N}+1))_{ch}
< 3/2+2|c|^2 \ , 
\label{5-31a}\\
& &2\sqrt{3}+2|c|^2
< ({\tilde f}({N}){\tilde f}({N}+1))_{sh}
< 7/2+2|c|^2 \ , 
\label{5-31b}
\end{eqnarray}
\end{subequations}
\vspace{-0.8cm}
\begin{subequations}\label{5-32}
\begin{eqnarray}
& &2+2|c|^2
< ({\tilde f}({N}){\tilde f}({N}+1))_{ch}
< 5/2+2|c|^2 \ , 
\label{5-32a}\\
& &\sqrt{6}+2|c|^2 
< ({\tilde f}({N}){\tilde f}({N}+1))_{sh}
< 5/2+2|c|^2 \ . 
\label{5-32b}
\end{eqnarray}
\end{subequations}
If $x=|c|^2$, the lower and upper limits of the inequalities 
(\ref{5-31}) and (\ref{5-32}) are the same as those shown in the 
inequality (\ref{5-27}) for $\lambda=1, 2$ and $\mu=3, 4$. 
Further, the differences of the upper and the lower limits 
are $3/2-\sqrt{2}=0.08579$, $7/2-2\sqrt{3}=0.03590$, 
$5/2-2=0.5$ and $5/2-\sqrt{6}=0.05051$, which are averagely rather small. 
Therefore, in high accuracy, we can put for the relation (\ref{5-25}), 
\begin{subequations}\label{5-33}
\begin{eqnarray}
& &({\tilde f}({N}){\tilde f}({N}+1))_{ch}
\simeq \sqrt{(2|c|^2+1)(2|c|^2+2)} \ , 
\label{5-33a}\\
& &({\tilde f}({N}){\tilde f}({N}+1))_{sh}
\simeq \sqrt{(2|c|^2+3)(2|c|^2+4)} \ . 
\label{5-33b}
\end{eqnarray}
\end{subequations}
In the same manner as the above, for the relation (\ref{5-26}), we have 
\begin{subequations}\label{5-34}
\begin{eqnarray}
& &({\tilde f}({N}){\tilde f}({N}+1))_{ch}
\simeq \sqrt{(2|c|^2+1)(2|c|^2+4)} \ , 
\label{5-34a}\\
& &({\tilde f}({N}){\tilde f}({N}+1))_{sh}
\simeq \sqrt{(2|c|^2+2)(2|c|^2+3)} \ . 
\label{5-34b}
\end{eqnarray}
\end{subequations}
With the use of the relations (\ref{4-10}), (\ref{5-16}) and 
(\ref{5-33}), we have the following forms for the deformation 
(\ref{5-11}) : 
\begin{subequations}\label{5-35}
\begin{eqnarray}
& &(\tau_+)_{ch} \simeq c^* \sqrt{1/2+|c|^2} = \tau_+(1/4) \ , 
\nonumber\\
& &(\tau_-)_{ch} \simeq c \sqrt{1/2+|c|^2} = \tau_-(1/4) \ , 
\label{5-35a}\\
& &(\tau_+)_{sh} \simeq c^* \sqrt{3/2+|c|^2} = \tau_+(3/4) \ , \nonumber\\
& &(\tau_-)_{sh} \simeq c \sqrt{3/2+|c|^2} = \tau_-(3/4) \ . 
\label{5-35b}
\end{eqnarray}
\end{subequations}
For the deformation (\ref{5-20}), the relations (\ref{4-10}), (\ref{5-24}) 
and (\ref{5-34}) give us the same expressions as those shown in the relation 
(\ref{5-35}). Judging from the property of the inequality (\ref{5-27}) 
already mentioned, the approximate expression (\ref{5-35}) may be valid 
in high 
accuracy. Therefore, we can conclude that $(\tau_{\pm,0})_{ch}$ and 
$(\tau_{\pm,0})_{sh}$ satisfy the relation of the Poisson bracket of 
the $su(1,1)$-generators in high accuracy. 
Further, the relation (\ref{5-13}) tells us that in this case, also, 
the expectation values of the products of $({\hat \tau}_{\pm,0})$ 
are not equal to the corresponding products of $(\tau_{\pm,0})_{ch}$ 
and $(\tau_{\pm,0})_{sh}$. 
%\begin{subequations}\label{5-35}
%\begin{eqnarray}
%& &|(\tau_\pm)_{ch}| \simeq |c| \sqrt{1/2+|c|^2} = |\tau_\pm(1/4)| \ , 
%\label{5-35a}\\
%& &|(\tau_\pm)_{sh}| \simeq |c| \sqrt{3/2+|c|^2} = |\tau_\pm(3/4)| \ . 
%\label{5-35b}
%\end{eqnarray}
%\end{subequations}
However, there exists another example which is in the situation 
similar to the above one. 
This example is discussed in \S 7.

\section{Deformation satisfying the relation of the Poisson bracket of the 
$su(1,1)$-algebra}

Until the present stage, we have discussed the deformations related to 
the boson-pair in connection to the $su(1,1)$-algebra. 
However, they do not satisfy the 
$su(1,1)$-algebra exactly. In this section, we show the deformation 
leading to this algebra exactly. 

We start in the following deformation : 
\begin{equation}\label{6-1}
{\tilde f}(2n)=2n+1 \ , \qquad {\tilde f}(2n-1)=1 \ . 
\end{equation}
In this case, $\dket{ch}$ and $\dket{sh}$ in (\ref{4-4}) 
can be written as 
\begin{subequations}\label{6-2}
\begin{eqnarray}
& &\dket{ch}=\sum_{n=0}^{\infty}\frac{\gamma^{2n}}{\sqrt{(2n)!}}(2n-1)!!
\ket{2n}=\exp(\gamma^2{\hat c}^{*2}/2) \ket{0} \ , 
\label{6-2a}\\
& &\dket{sh}=\sum_{n=0}^{\infty}\frac{\gamma^{2n+1}}{\sqrt{(2n+1)!}}(2n+1)!!
\ket{2n+1}=\gamma {\hat c}^*\exp(\gamma^2{\hat c}^{*2}/2) \ket{0} \ , 
\label{6-2b}
\end{eqnarray}
\end{subequations}
\vspace{-0.4cm}
\begin{subequations}\label{6-3}
\begin{eqnarray}
& &\Gamma_{ch}=\sum_{n=0}^{\infty}\frac{(|\gamma|^2)^{2n}}{(2n)!}[(2n-1)!!]^2
=\left(\sqrt{1-|\gamma|^4}\right)^{-1} \ , 
\label{6-3a}\\
& &\Gamma_{sh}=\sum_{n=0}^{\infty}\frac{(|\gamma|^2)^{2n+1}}{(2n+1)!}
[(2n+1)!!]^2
=|\gamma|^2\left(\sqrt{1-|\gamma|^4}\right)^{-3} \ . 
\label{6-3b}
\end{eqnarray}
\end{subequations}
The relation between $\dket{ch}$ and $\dket{sh}$ is as follows : 
\begin{equation}\label{6-4}
\dket{sh}=\gamma{\hat c}^*\dket{ch} \ , \qquad
\dket{ch}=\gamma^{-1}({\hat N}+1)^{-1}{\hat c}\dket{sh} \ .
\end{equation}
The states $\ket{ch}$ and $\ket{sh}$ obey the relation 
\begin{subequations}\label{6-5}
\begin{eqnarray}
& &(2{\hat \tau}_0+1/2)^{-1}{\hat \tau}_- \ket{ch}
=(\gamma^2/2)\ket{ch} \ , 
\label{6-5a}\\
& &(2{\hat \tau}_0+5/2)^{-1}{\hat \tau}_- \ket{sh}
=(\gamma^2/2)\ket{sh} \ . 
\label{6-5b}
\end{eqnarray}
\end{subequations}
It may be important to note that $\dket{ch}$ and $\dket{sh}$ are expressed 
in terms of ${\hat c}^{*2}/2$, which is a generator of the $su(1,1)$-algebra 
with $t=1/4$ and $3/4$, i.e., ${\hat \tau}_+$. 
The quantities $F_{ch}(|\gamma|^2)$ and $F_{sh}(|\gamma|^2)$ 
in (\ref{4-13}) are given 
in the form 
\begin{subequations}\label{6-6}
\begin{eqnarray}
& &F_{ch}(|\gamma|^2)=1-|\gamma|^4 \ , 
\label{6-6a}\\
& &F_{sh}(|\gamma|^2)=(1/3)\cdot(1-|\gamma|^4) \ . 
\label{6-6b}
\end{eqnarray}
\end{subequations}
With the use of the relation (\ref{4-14}), $|\gamma|^4$ can be expressed as 
follows : 
\begin{subequations}\label{6-7}
\begin{eqnarray}
& &|\gamma|^4=2|c|^2\cdot (1+2|c|^2)^{-1} \ , 
\label{6-7a}\\
& &|\gamma|^4=2|c|^2\cdot (3+2|c|^2)^{-1} \ . 
\label{6-7b}
\end{eqnarray}
\end{subequations}
Then, $F_{ch}(|\gamma|^2)$ and $F_{sh}(|\gamma|^2)$ are expressed as 
\begin{subequations}\label{6-8}
\begin{eqnarray}
& &F_{ch}(|\gamma|^2)=(1+2|c|^2)^{-1} \ , 
\label{6-8a}\\
& &F_{sh}(|\gamma|^2)=(3+2|c|^2)^{-1} \ . 
\label{6-8b}
\end{eqnarray}
\end{subequations}
%The relation (\ref{4-13}) can be given in the present case as follows : 
From the relation (\ref{4-13}), the following relations can be given 
in the present case : 
\begin{subequations}\label{6-9}
\begin{eqnarray}
& &\gamma=\sqrt{\sqrt{2}c}\cdot (1+2|c|^2)^{-1/4} \ , 
\quad 
\gamma^*=\sqrt{\sqrt{2}c^*}\cdot (1+2|c|^2)^{-1/4} \ , 
\label{6-9a}\\
& &\gamma=\sqrt{\sqrt{2}c}\cdot (3+2|c|^2)^{-1/4} \ , 
\quad 
\gamma^*=\sqrt{\sqrt{2}c^*}\cdot (3+2|c|^2)^{-1/4} \ . 
\label{6-9b}
\end{eqnarray}
\end{subequations}
%In order to obtain $(\tau_\pm)_{ch}$ and $(\tau_\pm)_{sh}$, it is 
%necessary to calculate 
The quantities 
$({\tilde f}({N}){\tilde f}({N}+1))_{ch}$ and 
$({\tilde f}({N}){\tilde f}({N}+1))_{sh}$ in the present case are 
calculated in the form 
\begin{subequations}\label{6-10}
\begin{eqnarray}
& &({\tilde f}({N}){\tilde f}({N}+1))_{ch}
=(1-|\gamma|^4)^{-1}=1+2|c|^2 \ , 
\label{6-10a}\\
& &({\tilde f}({N}){\tilde f}({N}+1))_{sh}
=3(1-|\gamma|^4)^{-1}=3+2|c|^2 \ . 
\label{6-10b}
\end{eqnarray}
\end{subequations}
Therefore, the relations (\ref{4-10a}) and (\ref{4-10b}) are written as 
\begin{subequations}\label{6-11}
\begin{eqnarray}
& &(\tau_+)_{ch}=c^*\sqrt{1/2+|c|^2}=\tau_+(1/4) \ , \nonumber\\
& &(\tau_-)_{ch}=c\sqrt{1/2+|c|^2}=\tau_-(1/4) \ , 
\label{6-11a}\\
& &(\tau_+)_{sh}=c^*\sqrt{3/2+|c|^2}=\tau_+(3/4) \ , \nonumber\\
& &(\tau_-)_{sh}=c\sqrt{3/2+|c|^2}=\tau_-(3/4) \ . 
\label{6-11b}
\end{eqnarray}
\end{subequations}
Thus, we can understand that the deformation (\ref{6-1}) satisfies 
the relation of the Poisson bracket of the $su(1,1)$-algebra exactly.

Next, we investigate the following deformation : 
\begin{equation}\label{6-12}
{\tilde f}(2n)=1 \ , \qquad {\tilde f}(2n-1)=2n-1 \ . 
\end{equation}
In this case, we have 
\begin{subequations}\label{6-13}
\begin{eqnarray}
\dket{ch}&=&\sum_{n=0}^{\infty}\frac{\gamma^{2n}}{\sqrt{(2n)!}}(2n-1)!!
\ket{2n}=\exp(\gamma^2{\hat c}^{*2}/2) \ket{0} \ , 
\label{6-13a}\\
\dket{sh}&=&\sum_{n=0}^{\infty}\frac{\gamma^{2n+1}}{\sqrt{(2n+1)!}}(2n-1)!!
\ket{2n+1} \nonumber\\
&=&\gamma {\hat c}^*({\hat N}+1)^{-1}
\exp(\gamma^2{\hat c}^{*2}/2) \ket{0} \ , 
\label{6-13b}
\end{eqnarray}
\end{subequations}
\vspace{-0.4cm}
\begin{subequations}\label{6-14}
\begin{eqnarray}
& &\Gamma_{ch}=\left(\sqrt{1-|\gamma|^4}\right)^{-1} \ , 
\label{6-14a}\\
& &\Gamma_{sh}=\sin^{-1}|\gamma|^2 \ . 
\label{6-14b}
\end{eqnarray}
\end{subequations}
The relation between $\dket{ch}$ and $\dket{sh}$ is as follows : 
\begin{equation}\label{6-15}
\dket{sh}=\gamma{\hat c}^*({\hat N}+1)^{-1}\dket{ch} \ , \qquad
\dket{ch}=\gamma^{-1}{\hat c}\dket{sh} \ .
\end{equation}
The states $\ket{ch}$ and $\ket{sh}$ satisfy 
\begin{subequations}\label{6-16}
\begin{eqnarray}
& &(2{\hat \tau}_0+1/2)^{-1}{\hat \tau}_-\ket{ch}
=(\gamma^2/2)\ket{ch} \ , 
\label{6-16a}\\
& &{\hat P}_0(2{\hat \tau}_0-1/2)^{-1}{\hat \tau}_-\ket{sh}
=(\gamma^2/2)\ket{sh} \ . 
\label{6-16b}
\end{eqnarray}
\end{subequations}
Here, ${\hat P}_0$ denotes a projection operator which gives 
us ${\hat P}_0\ket{2n}=0$ and ${\hat P}_0\ket{2n+1}=\ket{2n+1}$. 
It may be interesting to see that the state $\dket{ch}$ in the form 
(\ref{6-13a}) is nothing but the state $\dket{ch}$ shown in the relation 
(\ref{6-2a}), i.e., $t=1/4$. 
However, the state $\dket{sh}$, the partner of which is $\dket{ch}$, 
is different from the form shown in the 
relation (\ref{6-2b}). Concerning the above case, it may be interesting to 
investigate the following case :
\begin{equation}\label{6-17}
{\tilde f}(2n)=1 \ , \qquad {\tilde f}(2n-1)=2n+1 \ . 
\end{equation}
In this case, $\dket{ch}$ and $\dket{sh}$ can be expressed as 
\begin{subequations}\label{6-18}
\begin{eqnarray}
& &\dket{ch}=\sum_{n=0}^{\infty}\frac{\gamma^{2n}}{\sqrt{(2n)!}}(2n+1)!!
\ket{2n}=({\hat N}+1) \exp(\gamma^2{\hat c}^{*2}/2) \ket{0} \ , 
\label{6-18a}\\
& &\dket{sh}=\sum_{n=0}^{\infty}\frac{\gamma^{2n+1}}{\sqrt{(2n+1)!}}(2n+1)!!
\ket{2n+1}=\gamma {\hat c}^*\exp(\gamma^2{\hat c}^{*2}/2) \ket{0} \ , 
\label{6-18b}
\end{eqnarray}
\end{subequations}
\vspace{-0.4cm}
\begin{subequations}\label{6-19}
\begin{eqnarray}
& &\Gamma_{ch}=\left(\sqrt{1-|\gamma|^4}\right)^{-3} \ , 
\label{6-19a}\\
& &\Gamma_{sh}=|\gamma|^2\left(\sqrt{1-|\gamma|^4}\right)^{-3} \ . 
\label{6-19b}
\end{eqnarray}
\end{subequations}
Two states are related to each other through the relation 
\begin{equation}\label{6-20}
\dket{sh}=\gamma{\hat c}^*\left({\hat N}+1\right)^{-1}\dket{ch} \ , \qquad
\dket{ch}=\gamma^{-1}{\hat c}\dket{sh} \ .
\end{equation}
The above is the same as that given in the relation (\ref{6-15}). 
We know that the state $\dket{sh}$ in the form (\ref{6-18b}) is nothing but 
the state $\dket{sh}$ shown in the relation (\ref{6-2b}), i.e., 
$t=3/4$. However, 
the partner $\dket{ch}$ is different from the state 
$\dket{ch}$ shown in the relation (\ref{6-2a}). 
The two states $\ket{ch}$ and $\ket{sh}$ satisfy 
\begin{subequations}\label{6-21}
\begin{eqnarray}
& &(2{\hat \tau}_0+5/2)^{-1}{\hat \tau}_- \ket{ch}=(\gamma^2/2)\ket{ch} \ , 
\label{6-21a}\\
& &(2{\hat \tau}_0+5/2)^{-1}{\hat \tau}_- \ket{sh}=(\gamma^2/2)\ket{sh} \ . 
\label{6-21b}
\end{eqnarray}
\end{subequations}

Finally, we show two examples in which the situations are 
in the same as those given in the cases (\ref{6-1}) and (\ref{6-17}). 
These are two concrete examples of the deformation (\ref{5-3}). 
One is as follows : 
\begin{equation}\label{6-22}
{\tilde f}(2n)=\sqrt{2n+1} \ , \qquad {\tilde f}(2n-1)=\sqrt{2n-1} \ . 
\end{equation}
This case leads us to 
\begin{subequations}\label{6-23}
\begin{eqnarray}
\dket{ch}&=&\sum_{n=0}^{\infty}\frac{\gamma^{2n}}{\sqrt{(2n)!}}(2n-1)!!
\ket{2n}=\exp(\gamma^2{\hat c}^{*2}/2) \ket{0} \ , 
\label{6-23a}\\
\dket{sh}&=&\sum_{n=0}^{\infty}\frac{\gamma^{2n+1}}{\sqrt{(2n+1)!}}(2n+1)!!
(\sqrt{2n+1})^{-1}\ket{2n+1} \nonumber\\
&=&\gamma {\hat c}^*\left(\sqrt{{\hat N}+1}\right)^{-1}
\exp(\gamma^2{\hat c}^{*2}/2) \ket{0} \ , 
\label{6-23b}
\end{eqnarray}
\end{subequations}
\vspace{-0.4cm}
\begin{subequations}\label{6-24}
\begin{eqnarray}
& &\Gamma_{ch}=\left(\sqrt{1-|\gamma|^4}\right)^{-1} \ , 
\label{6-24a}\\
& &\Gamma_{sh}=|\gamma|^2\left(\sqrt{1-|\gamma|^4}\right)^{-1} \ . 
\label{6-24b}
\end{eqnarray}
\end{subequations}
%These two states are related to 
%\begin{equation}\label{6-22}
%\dket{sh}=\gamma{\hat c}^*\left(\sqrt{{\hat N}+1}\right)^{-1}\dket{ch} 
%\ , \qquad
%\dket{ch}=\gamma^{-1}\left(\sqrt{{\hat N}+1}\right)^{-1}{\hat c}\dket{sh} \ .
%\end{equation}
The relation between states is shown in the relation (\ref{5-4}). 
Clearly, $\dket{ch}$ is the state with $t=1/4$. However, 
the partner $\dket{sh}$ 
is different from the state with $t=3/4$. 
Second is given in the form 
\begin{equation}\label{6-25}
{\tilde f}(2n)=\sqrt{2n+1} \ , \qquad {\tilde f}(2n-1)=\sqrt{2n+1} \ . 
\end{equation}
This case gives us two states : 
\begin{subequations}\label{6-26}
\begin{eqnarray}
& &\dket{ch}=\sum_{n=0}^{\infty}\frac{\gamma^{2n}}{\sqrt{(2n)!}}(2n-1)!!
\sqrt{2n+1}\ket{2n}
=\sqrt{{\hat N}+1} \exp(\gamma^2{\hat c}^{*2}/2) \ket{0} \ , \qquad
\label{6-26a}\\
& &\dket{sh}=\sum_{n=0}^{\infty}\frac{\gamma^{2n+1}}{\sqrt{(2n+1)!}}(2n+1)!!
\ket{2n+1}
=\gamma {\hat c}^* \exp(\gamma^2{\hat c}^{*2}/2) \ket{0} \ , 
\label{6-26b}
\end{eqnarray}
\end{subequations}
\vspace{-0.4cm}
\begin{subequations}\label{6-27}
\begin{eqnarray}
& &\Gamma_{ch}=\left(\sqrt{1-|\gamma|^4}\right)^{-3} \ , 
\label{6-27a}\\
& &\Gamma_{sh}=|\gamma|^2 \left(\sqrt{1-|\gamma|^4}\right)^{-3} \ . 
\label{6-27b}
\end{eqnarray}
\end{subequations}
These two states are connected to each other through the relation 
(\ref{5-4}). 
%\begin{equation}\label{6-25}
%\dket{sh}=\gamma{\hat c}^*\left(\sqrt{{\hat N}+1}\right)^{-1}\dket{ch} \ , 
%\qquad
%\dket{ch}=\gamma^{-1}\left(\sqrt{{\hat N}+1}\right)^{-1}{\hat c}\dket{sh} \ .
%\end{equation}
It may be self-evident that $\dket{sh}$ is the state with $t=3/4$, but, 
the partner $\dket{ch}$ is different from the state with $t=1/4$. 
%The above two deformations are the examples of the deformation 
%(\ref{5-3}). 
The states shown in the relations (\ref{6-18a}) and (\ref{6-18b}) satisfy 
\begin{subequations}\label{6-28}
\begin{eqnarray}
& &(2{\hat \tau}_0+1/2)^{-1}{\hat \tau}_- \ket{ch}
=(\gamma^2/2)\ket{ch} \ , 
\label{6-28a}\\
& &{\hat P}_0\left(\sqrt{(2{\hat \tau}_0-1/2)(2{\hat \tau}_0+3/2)}
\right)^{-1}{\hat \tau}_- \ket{sh}
=(\gamma^2/2)\ket{sh} \ . 
\label{6-28b}
\end{eqnarray}
\end{subequations}
The states (\ref{6-23a}) and (\ref{6-23b}) obey 
\begin{subequations}\label{6-29}
\begin{eqnarray}
& &\left(\sqrt{(2{\hat \tau}_0+1/2)(2{\hat \tau}_0+5/2)}
\right)^{-1}{\hat \tau}_- \ket{ch}
=(\gamma^2/2)\ket{ch} \ , 
\label{6-29a}\\
& &(2{\hat \tau}_0+5/2)^{-1}{\hat \tau}_- \ket{sh}
=(\gamma^2/2)\ket{sh} \ . 
\label{6-29b}
\end{eqnarray}
\end{subequations}

\section{A possible deformation satisfying approximately the relation of the 
Poisson bracket of the $su(1,1)$-algebra}

In \S 6, we showed the deformation satisfying the relation of the Poisson 
bracket of the $su(1,1)$-algebra. In \S 5, the approximated form was 
discussed. In this section, again, we discuss the case in which the 
deformation satisfies the relation of the Poisson bracket 
in a well approximated form. Later, the reason why we discuss this case again 
will be mentioned. This case starts in the deformation 
\begin{equation}\label{7-1}
{\tilde f}(2n)=\sqrt{2n+1} \ , \qquad {\tilde f}(2n-1)=1 \ .
\end{equation}
This is an example of the deformation (\ref{5-3}). The states 
$\dket{ch}$ and $\dket{sh}$ can be written down as 
\begin{subequations}\label{7-2}
\begin{eqnarray}
& &\dket{ch}=\ket{0}+\sum_{n=1}^{\infty} \frac{\gamma^{2n}}{\sqrt{(2n)!}}
\sqrt{(2n-1)!!}\ \ket{2n} \ , 
\label{7-2a}\\
& &\dket{sh}=\sum_{n=0}^{\infty} \frac{\gamma^{2n+1}}{\sqrt{(2n+1)!}}
\sqrt{(2n+1)!!}\ \ket{2n+1} \ , \qquad\quad
\label{7-2b}
\end{eqnarray}
\end{subequations}
\vspace{-0.3cm}
\begin{subequations}\label{7-3}
\begin{eqnarray}
& &\Gamma_{ch}=1+\sum_{n=1}^{\infty} \frac{(|\gamma|^2)^{2n}}{(2n)!}
{(2n-1)!!} =\exp (|\gamma|^4/2) \ , 
\label{7-3a}\\
& &\Gamma_{sh}=\sum_{n=0}^{\infty} \frac{(|\gamma|^2)^{2n+1}}{(2n+1)!}
{(2n+1)!!} =|\gamma|^2\exp(|\gamma|^4/2) \ . 
\label{7-3b}
\end{eqnarray}
\end{subequations}
The relation between $\dket{ch}$ and $\dket{sh}$ is given in the relation 
(\ref{5-4}). The states $\ket{ch}$ and $\ket{sh}$ obey 
\begin{subequations}\label{7-4}
\begin{eqnarray}
& &\left(\sqrt{4{\hat \tau}_0+1}\right)^{-1}{\hat \tau}_- \ket{ch}
=(\gamma^2/2)\ket{ch} \ , 
\label{7-4a}\\
& &\left(\sqrt{4{\hat \tau}_0+3}\right)^{-1}{\hat \tau}_- \ket{sh}
=(\gamma^2/2)\ket{sh} \ . 
\label{7-4b}
\end{eqnarray}
\end{subequations}
In this case, $F_{ch}(|\gamma|^2)$ and $F_{sh}(|\gamma|^2)$ are expressed 
in the form 
\begin{equation}\label{7-5}
F_{ch}(|\gamma|^2)=F_{sh}(|\gamma|^2)=1 \ . 
\end{equation}
Then, for both cases, $(\gamma , \gamma^*)$ is obtained exactly as 
\begin{equation}\label{7-6}
\gamma=\sqrt{\sqrt{2}c} \ , \qquad \gamma^*=\sqrt{\sqrt{2}c^*} \ . 
\end{equation}
The expectation values $({\tilde f}(N){\tilde f}(N+1))_{ch}$ and 
$({\tilde f}(N){\tilde f}(N+1))_{sh}$ are calculated as follows : 
\begin{subequations}\label{7-7}
\begin{eqnarray}
& &({\tilde f}(N){\tilde f}(N+1))_{ch} = \sum_{n=0}^{\infty}
\frac{(|\gamma|^4/2)^n}{n!} \sqrt{2n+1}\cdot e^{-|\gamma|^4/2} \ , 
\label{7-7a}\\
& &({\tilde f}(N){\tilde f}(N+1))_{sh} = \sum_{n=0}^{\infty}
\frac{(|\gamma|^4/2)^n}{n!} \sqrt{2n+3}\cdot e^{-|\gamma|^4/2} \ , 
\label{7-7b}
\end{eqnarray}
\end{subequations}
As is given in Appendix B, we have the following approximate expression : 
\begin{equation}\label{7-8}
\sum_{n=0}^{\infty} \frac{(|\gamma|^4/2)^n}{n!} \sqrt{4t+2n}\cdot
e^{-|\gamma|^4/2} 
\sim \sqrt{4t+|\gamma|^4} \ . 
\end{equation}
With the use of the relation (\ref{7-8}) in the form (\ref{7-7}), we obtain 
\begin{subequations}\label{7-9}
\begin{eqnarray}
& &(\tau_+)_{ch} \sim c^*\sqrt{1/2+|c|^2} = \tau_+(1/4) \ , 
\nonumber\\
& &(\tau_-)_{ch} \sim c\sqrt{1/2+|c|^2} = \tau_-(1/4) \ , 
\label{7-9a}\\
& &(\tau_+)_{sh} \sim c^*\sqrt{3/2+|c|^2} = \tau_+(3/4) \ , 
\nonumber\\
& &(\tau_-)_{sh} \sim c\sqrt{3/2+|c|^2} = \tau_-(3/4) \ . 
\label{7-9b}
\end{eqnarray}
\end{subequations}
Of course, we used the relation (\ref{7-6}). Thus, under the high accuracy 
mentioned in Appendix B, the deformation (\ref{7-1}) gives us the 
relation of the Poisson bracket of the $su(1,1)$-algebra. 
% in high accuracy. 

Our main interest in this section is to discuss the expectation values 
of ${\hat \tau}_+{\hat \tau}_-$, ${\hat \tau}_0{\hat \tau}_-$ and 
${\hat \tau}_-^2$ for the states $\ket{ch}$ and $\ket{sh}$ adopted in 
this section. 
For this aim, the following formula is useful : 
\begin{subequations}\label{7-10}
\begin{eqnarray}
& &(c^{*2}c^2)_{ch}=|\gamma|^4(1+|\gamma|^4) \ , \nonumber\\
& &(c^{*}c^3)_{ch}=\gamma^2|\gamma|^4\sum_{n=0}^{\infty}
\frac{(|\gamma|^4/2)^n}{n!}\sqrt{2n+3}\cdot e^{-|\gamma|^4/2} \ , \nonumber\\
& &(c^4)_{ch}=\gamma^4\sum_{n=0}^{\infty}
\frac{(|\gamma|^4/2)^n}{n!}\sqrt{(2n+1)(2n+3)}\cdot e^{-|\gamma|^4/2} \ , 
\label{7-10a}\\
& &(c^{*2}c^2)_{sh}=|\gamma|^4(3+|\gamma|^4) \ , \nonumber\\
& &(c^{*}c^3)_{sh}=\gamma^2\biggl[|\gamma|^4\sum_{n=0}^{\infty}
\frac{(|\gamma|^4/2)^n}{n!}\sqrt{2n+5}
%\nonumber\\
%& &\qquad\qquad\qquad\qquad\qquad\qquad
+\sum_{n=0}^{\infty}\frac{(|\gamma|^4/2)^n}{n!}\sqrt{2n+3} \biggl]
\cdot e^{-|\gamma|^4/2} \ , \nonumber\\
& &(c^4)_{sh}=\gamma^4\sum_{n=0}^{\infty}
\frac{(|\gamma|^4/2)^n}{n!}\sqrt{(2n+3)(2n+5)}\cdot e^{-|\gamma|^4/2} \ , 
\label{7-10b}
\end{eqnarray}
\end{subequations}
With the case of the formula (\ref{7-10}), $(\tau_+\tau_-)_{ch}$, 
$(\tau_0\tau_-)_{ch}$, $(\tau_-^2)_{ch}$, $(\tau_+\tau_-)_{sh}$, 
$(\tau_0\tau_-)_{sh}$ and $(\tau_-^2)_{sh}$ are calculated as 
\begin{subequations}\label{7-11}
\begin{eqnarray}
& &(\tau_+\tau_-)_{ch}=|c|^2(1/2+|c|^2) \ , \nonumber\\
& &(\tau_0\tau_-)_{ch}\sim c/4\cdot \left[
\sqrt{1/2+|c|^2}+4|c|^2\sqrt{3/2+|c|^2}\right] \ , \nonumber\\
& &(\tau_-^2)_{ch}\sim c^2\sqrt{(1/2+|c|^2)(3/2+|c|^2)} \ , 
\label{7-11a}\\
& &(\tau_+\tau_-)_{sh}=|c|^2(3/2+|c|^2) \ , \nonumber\\
& &(\tau_0\tau_-)_{sh}\sim c/4\cdot \left[
3\sqrt{3/2+|c|^2}+4|c|^2\sqrt{5/2+|c|^2}\right] \ , \nonumber\\
& &(\tau_-^2)_{sh}\sim c^2\sqrt{(3/2+|c|^2)(5/2+|c|^2)} \ , 
\label{7-11b}
\end{eqnarray}
\end{subequations}
Of course, we used the relations (\ref{5-27}) and (\ref{7-8}), 
together with the relation (\ref{7-6}).

As is clear from the relation (\ref{7-4}), the form (\ref{7-11}) cannot be 
expressed in terms of simple product of $(\tau_{\pm,0})_{ch}$ and 
$(\tau_{\pm,0})_{sh}$. However, the forms (\ref{7-9}) and (\ref{7-11}) 
give us the following relation : 
\begin{subequations}\label{7-12}
\begin{eqnarray}
& &(\tau_+\tau_-)_{ch}/(\tau_+)_{ch}(\tau_-)_{ch} \sim 1 \ , 
\nonumber\\
& &(\tau_0\tau_-)_{ch}/(\tau_0)_{ch}(\tau_-)_{ch} \sim 
\frac{1+4|c|^2\sqrt{(3/2+|c|^2)/(1/2+|c|^2)}}{1+4|c|^2} \ , 
\nonumber\\
& &(\tau_-^2)_{ch}/(\tau_-)_{ch}^2 \sim \sqrt{3}\cdot 
\sqrt{(1+(2/3)|c|^2)/(1+2|c|^2)} \ , 
\label{7-12a}\\
& &(\tau_+\tau_-)_{sh}/(\tau_+)_{sh}(\tau_-)_{sh} \sim 1 \ , 
\nonumber\\
& &(\tau_0\tau_-)_{sh}/(\tau_0)_{sh}(\tau_-)_{sh} \sim 
\frac{1+(4/3)|c|^2\sqrt{(5/2+|c|^2)/(3/2+|c|^2)}}{1+(4/3)|c|^2} \ , 
\nonumber\\
& &(\tau_-^2)_{sh}/(\tau_-)_{sh}^2 \sim \sqrt{5/3}\cdot 
\sqrt{(1+(2/5)|c|^2)/(1+(2/3)|c|^2)} \ . 
\label{7-12b}
\end{eqnarray}
\end{subequations}
On the contrary, we have the form for the case discussed in \S 6 as follows : 
\begin{subequations}\label{7-13}
\begin{eqnarray}
& &(\tau_+\tau_-)_{ch}/(\tau_+)_{ch}(\tau_-)_{ch} 
=\frac{1+6|c|^2}{1+2|c|^2} \ , 
\nonumber\\
& &(\tau_0\tau_-)_{ch}/(\tau_0)_{ch}(\tau_-)_{ch} 
=\frac{1+20|c|^2}{1+4|c|^2} \ , 
\nonumber\\
& &(\tau_-^2)_{ch}/(\tau_-)_{ch}^2 =12 \ , 
\label{7-13a}\\
& &(\tau_+\tau_-)_{sh}/(\tau_+)_{sh}(\tau_-)_{sh} 
=\frac{1+(5/3)|c|^2}{1+(2/3)|c|^2} \ , 
\nonumber\\
& &(\tau_0\tau_-)_{sh}/(\tau_0)_{sh}(\tau_-)_{sh} 
=\frac{1+(20/9)|c|^2}{1+(4/3)|c|^2} \ , 
\nonumber\\
& &(\tau_-^2)_{sh}/(\tau_-)_{sh}^2 =20/3 \ . 
\label{7-13b}
\end{eqnarray}
\end{subequations}
The relations (\ref{7-12}) and (\ref{7-13}) tell us the following feature : 
The expectation values of ${\hat \tau}_+{\hat \tau}_-$, 
${\hat \tau}_0{\hat \tau}_-$ and ${\hat \tau}_-^2$ 
are, of course, not equal to the products of the expectation values of 
${\hat \tau}_{\pm,0}$. However, compared with the case (\ref{7-13}), the case 
(\ref{7-12}) makes the expectation values of the products approach to the 
products of the expectation values. For example, the relation (\ref{7-12a}) 
shows us that, in the case of 
$(\tau_0\tau_-)_{ch}/(\tau_0)_{ch}(\tau_-)_{ch}$, at $|c|^2=0$ and 
$|c|^2\rightarrow \infty$, this ratio is equal to 1 and around $|c|^2=0.40$, 
this ratio takes the maximum value 1.2787. The case (\ref{7-12b}) is also 
in the same situation as the above. In this case, around $|c|^2=1.20$, 
the maximum value appears as 1.1050. 
The ratios $(\tau_-^2)_{ch}/(\tau_-)_{ch}^2$ and  
$(\tau_-^2)_{sh}/(\tau_-)_{sh}^2$ show the feature that at $|c|^2=0$ 
they have the maximum values $\sqrt{3}$ and $\sqrt{5/3}$, respectively, 
and gradually decrease to 1 at the limit $|c|^2\rightarrow \infty$. 
In the case of $\ket{ch^0}$ and $\ket{sh^0}$, the ratios defined in the 
relation (\ref{7-12}) or (\ref{7-13}) are exactly equal to unity 
in any region of $|c|^2$. In relation to the above statement, we contact 
with the states $\ket{ch}$ and $\ket{sh}$ in \S 5. This case also satisfies 
the relation of the Poisson bracket approximately. However, the ratios 
defined in the relation (\ref{7-12}) or (\ref{7-13}) are not equal to 1, 
for example, as shown in the following : 
\begin{subequations}\label{7-14}
\begin{eqnarray}
& &(\tau_+\tau_-)_{ch}/(\tau_+)_{ch}(\tau_-)_{ch} 
\sim \frac{1+4|c|^2}{1+2|c|^2} \ , 
\label{7-14a}\\
& &(\tau_+\tau_-)_{sh}/(\tau_+)_{sh}(\tau_-)_{sh} 
\sim \frac{1+|c|^2}{1+(2/3)|c|^2} \ . 
\label{7-14b}
\end{eqnarray}
\end{subequations}
The relation (\ref{7-14}) suggests us that the states $\ket{ch}$ and 
$\ket{sh}$ for the ratio (\ref{7-14}) are in the intermediate 
situation between the states $\ket{ch}$ and $\ket{sh}$ for the ratios 
(\ref{7-12}) and (\ref{7-13}). This is a reply of the promise mentioned 
in the final part of \S 5.

The above analysis gives us a suggestion mentioned below. 
The set $(\tau_{\pm,0}(t); t=1/4,3/4)$ shown in the relation (\ref{3-12}) 
denotes a classical counterpart of the $su(1,1)$-algebra defined in 
the relation (\ref{2-3}). 
We know the boson-pair coherent state which leads us to the classical 
counterpart (\ref{3-12}) through the expectation values of 
$({\hat \tau}_{\pm,0})$, i.e., the states (\ref{6-2a}) and (\ref{6-2b}). 
The expectation values are parametrized in terms of a quite simple 
function for the variable $(c, c^*)$. 
Therefore, the deviation of the expectation values of the products of 
$({\hat \tau}_{\pm,0})$ from the corresponding products of the 
expectation values of $({\hat \tau}_{\pm,0})$, means a kind of quantal 
fluctuation and it seems to be rather large. 
On the other hand, the states (\ref{3-2a}) and (\ref{3-2b}) cannot lead us to 
the relation (\ref{3-12}) and the expectation values parametrized in terms of 
$(c, c^*)$ are of the complicated form. 
However, concerning the products of $({\hat \tau}_{\pm,0})$, the forms are 
simply produced. This means that, at the stage of calculating the expectation 
values of $({\hat \tau}_{\pm,0})$, the fluctuation is taken into account 
and it is rather large. 
The states (\ref{7-2a}) and (\ref{7-2b}) are in the intermediate 
situation. The fluctuation contained in the expectation values of 
$({\hat \tau}_{\pm,0})$ is not so large and the fluctuation contained in 
the product is also not so large. 
Therefore, further comparative investigation may be an interesting 
problem.

%\section{Formalism in the boson mapping method}
\section{Deformation in the boson mapping method}

In this paper, we have investigated various deformations for the boson-pair 
coherent state. Characteristic point discussed in this paper can be found 
in stressing the even-odd boson number difference. As was mentioned in \S 1, 
our treatment\cite{two,three,four} was performed 
by using 
the MYT boson mapping method. 
However, in this paper, we treated the deformation 
in the original boson space. 

As a final discussion, we mention the relation of the present treatment 
to the boson mapping. For this aim, we prepare another boson space 
constructed by the boson $({\hat d} , {\hat d}^*)$. 
The orthogonal set is given by $\{ \rket{n} \}$ : 
\begin{equation}\label{8-1}
\rket{n}=(1/\sqrt{n!})\cdot {\hat d}^{*n} \rket{0} \ , \qquad
({\hat d}\rket{0}=0) \ .
\end{equation}
The mapping operator can be defined as 
\begin{subequations}\label{8-2}
\begin{eqnarray}
& &{\hat U}^{(e)} = \sum_{n=0}^{\infty} \rket{n}\bra{2n} \ , 
\label{8-2a}\\
& &{\hat U}^{(o)} = \sum_{n=0}^{\infty} \rket{n}\bra{2n+1} \ . 
\label{8-2b}
\end{eqnarray}
\end{subequations}
Here,it should be noted that the boson image of $\ket{2n}$ is the same as 
that of $\ket{2n+1}$, i.e., $\rket{n}$. 
In this paper, the even and the odd boson number states are treated 
separately and, then, the confusion does not happen. 
The operators $({\hat \tau}_{\pm,0})$ are mapped on the following 
forms : 
\begin{subequations}\label{8-3}
\begin{eqnarray}
& &{\tilde \tau}_{\pm,0}(e)={\hat U}^{(e)}{\hat \tau}_{\pm,0}
{\hat U}^{(e)\dagger} \ , 
\label{8-3a}\\
& &{\tilde \tau}_{\pm,0}(o)={\hat U}^{(o)}{\hat \tau}_{\pm,0}
{\hat U}^{(o)\dagger} \ , 
\label{8-3b}
\end{eqnarray}
\end{subequations}
\vspace{-0.5cm}
\begin{subequations}\label{8-4}
\begin{eqnarray}
& &{\tilde \tau}_+(e)={\hat d}^*\sqrt{1/2+{\hat d}^*{\hat d}} \ , 
\qquad
{\tilde \tau}_-(e)=\sqrt{1/2+{\hat d}^*{\hat d}} \ {\hat d}\ , 
\nonumber\\
& &{\tilde \tau}_0(e)={\hat d}^*{\hat d} +1/4 \ ,
\label{8-4a}\\
& &{\tilde \tau}_+(o)={\hat d}^*\sqrt{3/2+{\hat d}^*{\hat d}} \ , 
\qquad
{\tilde \tau}_-(o)=\sqrt{3/2+{\hat d}^*{\hat d}} \ {\hat d}\ , 
\nonumber\\
& &{\tilde \tau}_0(o)={\hat d}^*{\hat d} +3/4 \ . 
\label{8-4b}
\end{eqnarray}
\end{subequations}
The form (\ref{8-4}) should be compared with the form (\ref{3-12}).

The images of $\ket{ch}$ and $\ket{sh}$, which we denote $\rket{ch}$ and 
$\rket{sh}$, respectively, are given in the form 
\begin{subequations}\label{8-5}
\begin{eqnarray}
& &\rket{ch}={\hat U}^{(e)} \ket{ch}
=\left(\sqrt{\Gamma_{ch}}\right)^{-1} \rdket{ch} \ , 
\label{8-5a}\\
& &\rket{sh}={\hat U}^{(o)} \ket{sh}
=\left(\sqrt{\Gamma_{sh}}\right)^{-1} \rdket{sh} \ , \qquad\qquad
\label{8-5b}
\end{eqnarray}
\end{subequations}
\vspace{-0.3cm}
\begin{subequations}\label{8-6}
\begin{eqnarray}
& &\rdket{ch}=\rket{0}+\sum_{n=1}^{\infty}
\frac{\gamma^{2n}}{\sqrt{(2n)!}} {\tilde f}(0)\cdots {\tilde f}(2n-1)
\rket{n} \ , 
\label{8-6a}\\
& &\rdket{sh}=\sum_{n=0}^{\infty}
\frac{\gamma^{2n+1}}{\sqrt{(2n+1)!}} {\tilde f}(0)\cdots {\tilde f}(2n)
\rket{n} \ . 
\label{8-6b}
\end{eqnarray}
\end{subequations}
The states $\rdket{ch}$ and $\rdket{sh}$ can be rewritten as follows : 
\begin{subequations}\label{8-7}
\begin{eqnarray}
& &\rdket{ch}=\exp\left[\frac{\gamma^2}{\sqrt{2}}{\hat d}^*
\frac{{\tilde f}(2{\hat M})
{\tilde f}(2{\hat M}+1)}{\sqrt{2{\hat M}+1}}\right] \rket{0} \ , 
\label{8-7a}\\
& &\rdket{sh}=\gamma \exp\left[\frac{\gamma^2}{\sqrt{2}}
{\hat d}^*\frac{{\tilde f}(2{\hat M}+1)
{\tilde f}(2{\hat M}+2)}{\sqrt{2{\hat M}+3}}\right] \rket{0} \ , 
\label{8-7b}
\end{eqnarray}
\end{subequations}
\vspace{-0.3cm}
\begin{equation}\label{8-8}
{\hat M}={\hat d}^*{\hat d} \ . \qquad\qquad\qquad\qquad\qquad
\end{equation}
It is interesting to see that $\rdket{ch}$ and $\rdket{sh}$ are 
deformed from the conventional boson coherent state 
$\exp((\gamma^2/\sqrt{2})\cdot{\hat d}^*)\rket{0}$.

Let us discuss three typical examples. First is as follows : 
\begin{equation}\label{8-9}
{\tilde f}(2n)=1 \ , \qquad {\tilde f}(2n-1)=1 \ . 
\end{equation}
This case corresponds to $\dket{ch^0}$ and $\dket{sh^0}$ treated 
in \S 2. In this case, we have 
\begin{subequations}\label{8-10}
\begin{eqnarray}
& &\rdket{ch}=\exp\left[(\gamma^2/2){\hat d}^*(\sqrt{1/2+{\hat M}})^{-1} 
\right] \rket{0} \ , 
\label{8-10a}\\
& &\rdket{sh}=\gamma\exp\left[(\gamma^2/2){\hat d}^*(\sqrt{3/2+{\hat M}})^{-1} 
\right] \rket{0} \ . 
\label{8-10b}
\end{eqnarray}
\end{subequations}
Second is the case discussed in \S 6, i.e., 
\begin{equation}\label{8-11}
{\tilde f}(2n)=2n+1 \ , \qquad {\tilde f}(2n-1)=1 \ . 
\end{equation}
The above gives us 
\begin{subequations}\label{8-12}
\begin{eqnarray}
& &\rdket{ch}=\exp\left[\gamma^2{\hat d}^* \sqrt{1/2+{\hat M}} 
\right] \rket{0} \ , 
\label{8-12a}\\
& &\rdket{sh}=\gamma\exp\left[\gamma^2{\hat d}^* \sqrt{3/2+{\hat M}} 
\right] \rket{0} \ . 
\label{8-12b}
\end{eqnarray}
\end{subequations}
Third case corresponds to the deformation (\ref{7-1}), i.e., 
\begin{equation}\label{8-13}
{\tilde f}(2n)=\sqrt{2n+1} \ , \qquad {\tilde f}(2n-1)=1 \ . 
\end{equation}
We have 
\begin{subequations}\label{8-14}
\begin{eqnarray}
& &\rdket{ch}=\exp\left(\gamma^2/\sqrt{2}\cdot{\hat d}^* \right) \rket{0} \ , 
\label{8-14a}\\
& &\rdket{sh}=\gamma\exp\left(\gamma^2/\sqrt{2}\cdot
{\hat d}^* \right) \rket{0} \ . 
\label{8-14b}
\end{eqnarray}
\end{subequations}

We can see that the states $\rdket{ch}$ and $\rdket{sh}$ shown in the 
relation (\ref{8-14}) are identical with the conventional boson 
coherent states and $\rdket{ch}$ is of the same form as $\rdket{sh}$. 
The operators, for example, $({\hat \tau}_{\pm,0})$ acting on the states 
$\rdket{ch}$ and $\rdket{sh}$ are different from each other. 
Therefore, the expectation values are different for each other. 
It may be self-evident that the states (\ref{8-10}) and (\ref{8-12}) 
are deformed from the states (\ref{8-14}) by the functions 
$(\sqrt{1/2+{\hat M}})^{-1}$, $(\sqrt{3/2+{\hat M}})^{-1}$, 
$\sqrt{1/2+{\hat M}}$ and $\sqrt{3/2+{\hat M}}$. 
The extended versions of these two cases are adopted in Ref.\citen{one}. 
In this sense, it may be interesting to investigate the extension 
of the state (\ref{8-14}) for the model adopted in Ref.\citen{one}.

In this paper, Part (I), we discussed the boson-pair coherent states with the 
even- and the odd-boson number separately, but, intimately. 
Therefore, one defect can be found in the point that the expectation 
value of ${\hat c}$ (or ${\hat c}^*$) itself cannot be calculated. 
In order to make it possible, we must unify the two states $\ket{ch}$ 
and $\ket{sh}$. Prototype of this work is found in Ref.\citen{nine}, 
where only a certain special case was discussed and, then, 
in (II), we will extend the idea mentioned in Ref.\citen{nine}.

%\section*{Acknowledgements}
%\vspace{0.2cm}
%
%Two of the authors (Y. T. and M. Y.) would like to express their thanks to 
%Professor J. da Provid\^encia, one of co-authors of this paper. 
%He invited them to Coimbra in summer of 2002. During this stay, this work 
%was initiated. 

\appendix
\section{The proof of the relations (\ref{3-11})}

The aim of this Appendix is to give the proof of the relations 
(\ref{3-11a}) and (\ref{3-11b}). In the region $|\gamma|^2 \sim 0$, 
$F_{ch^0}(|\gamma|^2)$ and $F_{sh^0}(|\gamma|^2)$ shown in the 
relations (\ref{3-7a}) and (\ref{3-7b}), respectively, are expressed as 
\begin{subequations}\label{A-1}
\begin{eqnarray}
& &F_{ch^0}(|\gamma|^2) \sim 1+(1/3)|\gamma|^4-(1/45)|\gamma|^8 \ , 
\label{A-1a}\\
& &F_{sh^0}(|\gamma|^2) \sim 3+(1/5)|\gamma|^4-(1/175)|\gamma|^8 \ . 
\label{A-1b}
\end{eqnarray}
\end{subequations}
On the other hand, the asymptotic forms in the limit $|\gamma|^2 \rightarrow 
\infty$ are given as 
\begin{subequations}\label{A-2}
\begin{eqnarray}
& &F_{ch^0}(|\gamma|^2) \longrightarrow |\gamma|^2 \ , 
\label{A-2a}\\
& &F_{sh^0}(|\gamma|^2) \longrightarrow |\gamma|^2 +1 \ . 
\label{A-2b}
\end{eqnarray}
\end{subequations}
Our present aim is to get possible approximate expressions of 
$F_{ch^0}(|\gamma|^2)$ and $F_{sh^0}(|\gamma|^2)$, which reproduce 
the above two limiting cases, in terms of $2|c|^2$. 
Concerning the case $|\gamma|^2 \sim 0$, the relation (\ref{3-9}) 
combined with the form shown on the right-hand side of the relation 
(\ref{A-1}) leads us to the expression 
\begin{subequations}\label{A-3}
\begin{eqnarray}
& &F_{ch^0}(|\gamma|^2) \sim 1+(1/3)\cdot 2|c|^2+(4/45)\cdot (2|c|^2)^2 \ , 
\label{A-3a}\\
& &F_{sh^0}(|\gamma|^2) \sim 3+(3/5)\cdot 2|c|^2+(12/175)\cdot (2|c|^2)^2 \ . 
\label{A-3b}
\end{eqnarray}
\end{subequations}
Under the initial condition $|\gamma|^4=2|c|^2$, the above expression 
is derived iteratively. Therefore, the above form is effective in the 
case $2|c|^2 \sim 0$. Under the same procedure as that in the case 
$|\gamma|^2 \sim 0$, we can derive the following expression for the 
case $|\gamma|^2 \rightarrow \infty$ : 
\begin{subequations}\label{A-4}
\begin{eqnarray}
& &F_{ch^0}(|\gamma|^2) \longrightarrow 2|c|^2 \ , 
\label{A-4a}\\
& &F_{sh^0}(|\gamma|^2) \longrightarrow 2|c|^2+2 \ . 
\label{A-4b}
\end{eqnarray}
\end{subequations}
Of course, the above form is useful for the case $2|c|^2 \rightarrow \infty$. 

With the use of the expressions (\ref{A-3}) and (\ref{A-4}), 
let us derive an unified approximate form which includes the above two 
limiting cases. First, we define the function $G_{ch^0}(2|c|^2)$ and 
$G_{sh^0}(2|c|^2)$ through the relation 
\begin{subequations}\label{A-5}
\begin{eqnarray}
& &F_{ch^0}(|\gamma|^2)=G_{ch^0}(2|c|^2)+2|c|^2 \ , 
\label{A-5a}\\
& &F_{sh^0}(|\gamma|^2)=G_{sh^0}(2|c|^2)+(2|c|^2+2) \ . 
\label{A-5b}
\end{eqnarray}
\end{subequations}
Then, $G_{ch^0}(2|c|^2)$ and $G_{sh^0}(2|c|^2)$ should obey the 
following form : 
\begin{eqnarray}
& &G_{ch^0}(2|c|^2) \sim 1-(2/3)\cdot 2|c|^2+(4/45)\cdot (2|c|^2)^2 \ , 
\nonumber\\
& &G_{sh^0}(2|c|^2) \sim 1-(2/5)\cdot 2|c|^2+(12/175)\cdot (2|c|^2)^2 \ , 
\ \ (\hbox{\rm for}\ 2|c|^2 \sim 0) 
\label{A-6}\\
& &G_{ch^0}(2|c|^2) \sim 0 \ , \nonumber\\
& &G_{sh^0}(2|c|^2) \sim 0 \ . \ \ (\hbox{\rm for}\ 2|c|^2 \rightarrow \infty) 
\label{A-7}
\end{eqnarray}
As a possible candidate which includes the both limiting cases unifyingly, 
we can find the following form : 
\begin{subequations}\label{A-8}
\begin{eqnarray}
G_{ch^0}(2|c|^2) &\sim& 
H_{ch^0}(2|c|^2) \nonumber\\
&=& \exp \left[-\left((2/3)\cdot 2|c|^2+(2/15)\cdot (2|c|^2)^2
\right)\right] \ , 
\label{A-8a}\\
G_{sh^0}(2|c|^2) &\sim& 
H_{sh^0}(2|c|^2) \nonumber\\
&=& \exp \left[-\left((2/5)\cdot 2|c|^2+(2/175)\cdot (2|c|^2)^2
\right)\right] \ . 
\label{A-8b}
\end{eqnarray}
\end{subequations}
The above form has been shown in the relation (\ref{3-11}). 
It is easily verified that the relation (\ref{A-8}) is reduced to the 
forms (\ref{A-6}) and (\ref{A-7}) at the conditions $2|c|^2 \sim 0$ 
and $2|c|^2 \rightarrow \infty$, respectively.

The validity of the approximate expressions (\ref{3-11a}) and (\ref{3-11b}) 
can be checked in a way mentioned below. The relation (\ref{3-9}) gives 
us the following form : 
\begin{subequations}\label{A-9}
\begin{eqnarray}
& &|\gamma|^2=\sqrt{2|c|^2\cdot[H_{ch^0}(2|c|^2)+2|c|^2]} \ , 
\label{A-9a}\\
& &|\gamma|^2=\sqrt{2|c|^2\cdot[H_{sh^0}(2|c|^2)+(2|c|^2+2)]} \ . 
\label{A-9b}
\end{eqnarray}
\end{subequations}
On the other hand, the relation (\ref{3-9}) can be rewritten as 
\begin{subequations}\label{A-10}
\begin{eqnarray}
& &2|c|^2=|\gamma|^4\cdot F_{ch^0}(|\gamma|^2)^{-1}
=|\gamma|^2 \tanh |\gamma|^2 \ , 
\label{A-10a}\\
& &2|c|^2=|\gamma|^4\cdot F_{sh^0}(|\gamma|^2)^{-1}
=|\gamma|^2 \coth |\gamma|^2 -1 \ . 
\label{A-10b}
\end{eqnarray}
\end{subequations}
Substituting various values of $2|c|^2$ (the initial values of 
$2|c|^2$) into the relation (\ref{A-9}), we have the value of 
$|\gamma|^2$ corresponding to each $2|c|^2$. Then, substituting the 
values of $|\gamma|^2$ into the relation (\ref{A-10}), we have the value of 
$2|c|^2$ corresponding to each $|\gamma|^2$ (the resultant values 
of $2|c|^2$). 
Then, if the resultant value of $2|c|^2$ coincides with the 
initial value in the region $0 \leq 2|c|^2 < \infty$ in high accuracy, 
the expression (\ref{3-11}) is well approximated. 
Tables Ia and Ib show the results based on the above procedure. 
We can see that the expression (\ref{3-11}) is rather well approximated.

\begin{table}[t]
%\begin{wraptable}{l}{\halftext}
\caption{\hbox{}\hspace{-0.25cm}a }
\label{table:Ia}
\begin{center}
\begin{tabular}{c|c|c} \hline \hline
initial value of $2|c|^2$ & $|\gamma|^2$ & resultant value of $2|c|^2$ 
\\ \hline 
0.10 & 0.3216 & 0.1000 \\
0.25 & 0.5219 & 0.2501 \\
0.50 & 0.7723 & 0.5007 \\
1.0 & 1.2039 & 1.0050 \\
2.0 & 2.0759 & 2.0116 \\
3.0 & 3.0203 & 3.0060 \\
4.0 & 4.0041 & 4.0015 \\
5.0 & 5.0006 & 5.0002 \\
6.0 & 6.0001 & 6.0000 \\
7.0 & 7.0000 & 7.0000 \\
\hline
\end{tabular}
\end{center}
%\end{wraptable}
\end{table}

\setcounter{table}{0}
\begin{table}[t]
%\begin{wraptable}{l}{\halftext}
\caption{\hbox{}\hspace{-0.25cm}b }
\label{table:Ib}
\begin{center}
\begin{tabular}{c|c|c} \hline \hline
initial value of $2|c|^2$ & $|\gamma|^2$ & resultant value of $2|c|^2$ 
\\ \hline 
0.10 & 0.5532 & 0.1000 \\
0.25 & 0.8880 & 0.2500 \\
0.50 & 1.2877 & 0.4999 \\
1.0 & 1.9138 & 0.9990 \\
2.0 & 2.9763 & 1.9918 \\
3.0 & 3.9768 & 2.9796 \\
4.0 & 4.9672 & 3.9676 \\
5.0 & 5.9589 & 4.9590 \\
6.0 & 6.9542 & 5.9542 \\
7.0 & 7.9526 & 6.9526 \\
8.0 & 8.9530 & 7.9530 \\
9.0 & 9.9548 & 8.9548 \\
10.0 & 10.9571 & 9.9571 \\
20.0 & 20.9762 & 19.9762 \\
50.0 & 50.9902 & 49.9902 \\
\hline
\end{tabular}
\end{center}
%\end{wraptable}
\end{table}

%\appendix
\section{
The interpretation of the relation (\ref{7-8})}

The interpretation of the relation (\ref{7-8}) starts in the following 
relations :
\begin{subequations}\label{B-1}
\begin{eqnarray}
& &\sqrt{x+1} \geq 1+(1/2)x-(1/2)(\sqrt{\epsilon +1}+1)^{-2} x^2 \ , 
\qquad (x \geq \epsilon) 
\label{B-1a}\\
& &\sqrt{x+1} < 1+(1/2)x-(1/2)(\sqrt{\epsilon +1}+1)^{-2} x^2 \ , 
\qquad (-1 \le x < \epsilon) 
\label{B-1b}
\end{eqnarray}
\end{subequations}
\vspace{-0.3cm}
\begin{eqnarray}
& &\sum_{n=0}^{n_0} \frac{n_0^n}{n!}e^{-n_0} \simeq C(n_0) \ , 
\quad
C(n_0)=\sqrt{\frac{\pi n_0+2}{4\pi n_0+2}}
\left(1+(10/21)\frac{\sqrt{n_0}}{n_0+1}\right) \ , 
\label{B-2}\\
& &\frac{n_0^{n_0}}{n_0 !}e^{-n_0} \simeq D(n_0) \ , \quad
D(n_0)=\frac{1}{\sqrt{2\pi n_0+1}}\left(
1-(1/10)\frac{\sqrt{n_0}}{(n_0+1)^4}\right) \ . 
\label{B-3}
\end{eqnarray}
The proof of the relation (\ref{B-1}) is performed through the identity 
\begin{eqnarray}\label{B-4}
& &\sqrt{x+1}-\left[ 1+(1/2)x-(1/2)(\sqrt{\epsilon+1}+1)^{-2}x^2\right]
\nonumber\\
&=&(1/2)x^2(x-\epsilon)(\sqrt{\epsilon+1}+\sqrt{x+1}+2)
(\sqrt{\epsilon+1}+1)^{-2}\nonumber\\
& &\times (\sqrt{x+1}+1)^{-2}
(\sqrt{\epsilon+1}+\sqrt{x+1})^{-1} \ .
\end{eqnarray}
Concerning the relation (\ref{B-2}) and (\ref{B-3}), 
they are trivial at $n_0=0$ and at $n_0 \rightarrow \infty$, 
they show well-known asymptotic behavior. 
The behaviors in the region $n_0=1,2,3,\cdots$ are checked numerically 
and they are shown in Tables IIa and IIb.

\setcounter{table}{1}
\begin{table}[t]
%\begin{wraptable}{l}{\halftext}
\parbox{\halftext}{
\caption{\hbox{}\hspace{-0.25cm}a }
\label{table:IIa}
\begin{center}
\begin{tabular}{c|c|c} \hline \hline
$n_0$ & $\sum_{n=0}^{n_0}n_0^n / n! \cdot e^{-n_0}$ & $C(n_0)$ 
\\ \hline 
1 & 0.7358 & 0.7356 \\
2 & 0.6768 & 0.6766 \\
3 & 0.6472 & 0.6471 \\
4 & 0.6288 & 0.6285 \\
5 & 0.6160 & 0.6154 \\
6 & 0.6063 & 0.6055 \\
7 & 0.5983 & 0.5977 \\
\hline
\end{tabular}
\end{center}
%\end{wraptable}
%\end{table}
%
}
\setcounter{table}{1}
%\begin{table}[t]
%\begin{wraptable}{l}{\halftext}
\parbox{\halftext}{
\caption{\hbox{}\hspace{-0.25cm}b }
\label{table:IIb}
\begin{center}
\begin{tabular}{c|c|c} \hline \hline
$n_0$ & $n_0^{n_0} / n_0 ! \cdot e^{-n_0}$ & $D(n_0)$ 
\\ \hline 
1 & 0.3679 & 0.3682 \\
2 & 0.2707 & 0.2710 \\
3 & 0.2240 & 0.2244 \\
4 & 0.1954 & 0.1956 \\
5 & 0.1755 & 0.1756 \\
6 & 0.1606 & 0.1607 \\
7 & 0.1490 & 0.1491 \\
\hline
\end{tabular}
\end{center}
%\end{wraptable}
}
\end{table}

First, we note two relations 
\begin{eqnarray}
& &\sqrt{n+2t}=\sqrt{n_0+2t}\sqrt{1+\frac{n-n_0}{n_0+2t}} \ , 
\label{B-5}\\
& &\sum_{n=0}^{\infty} n\cdot \frac{n_0^n}{n!}\cdot e^{-n_0}
=n_0 \ . 
\label{B-6}
\end{eqnarray}
Hereafter, we treat $n_0$ as a positive integer. 
For $n=n_0+1, n_0+2, \cdots$, the relations (\ref{B-1a}) and 
(\ref{B-5}) give us 
\begin{equation}\label{B-7}
\sqrt{n+2t} \geq f(n ; n_0 t) \ , \qquad
(n=n_0+1, n_0+2, \cdots)
\end{equation}
\vspace{-0.3cm}
\setcounter{equation}{6}
\begin{subequations}
\begin{eqnarray}
f(n;n_0t)&=&\sqrt{n_0+2t}\nonumber\\
& &\times \left[ 1+\frac{1}{2}\left(\frac{n-n_0}{n_0+2t}\right)
-\frac{1}{2}\left(\frac{\sqrt{n_0+2t}}{\sqrt{n_0+1+2t}+\sqrt{n_0+2t}}
\right)^2 \left(\frac{n-n_0}{n_0+2t}\right)^2\right] \ . 
\nonumber\\
& &
\label{B-7a}
\end{eqnarray}
\end{subequations}
Here, for the application of the relation (\ref{B-1a}), $x$ and $\epsilon$ 
are regarded as 
\begin{equation}\label{B-8}
x=\frac{n-n_0}{n_0+2t} \ , \qquad 
\epsilon=\frac{1}{n_0+2t} \ . 
\end{equation}
Of course, we have 
\begin{equation}\label{B-9}
\sqrt{n+2t} < f(n;n_0t) \ . \qquad
(n=0, 1, 2,\cdots , n_0)
\end{equation}
With the use of the relation (\ref{B-7}), the following inequality 
is derived : 
\begin{eqnarray}\label{B-10}
& &\sum_{n=0}^{\infty}\frac{n_0^n}{n!}\sqrt{n+2t}
=\sum_{n=n_0+1}^{\infty}\frac{n_0^n}{n!}\sqrt{n+2t}
+\sum_{n=0}^{n_0}\frac{n_0^n}{n!}\sqrt{n+2t} \nonumber\\
&>& \sum_{n=n_0+1}^{\infty}\frac{n_0^n}{n!}f(n;n_0t)
+\sum_{n=0}^{n_0}\frac{n_0^n}{n!}\sqrt{n+2t} \nonumber\\
&=& F_1 + F_2 \ , 
\end{eqnarray}
\vspace{-0.3cm}
\setcounter{equation}{9}
\begin{subequations}
\begin{eqnarray}
& &F_1=\sum_{n=0}^{\infty}\frac{n_0^n}{n!}f(n;n_0t) \ , 
\label{B-10a}\\
& &F_2=-\sum_{n=0}^{n_0}\frac{n_0^n}{n!}
\left(f(n;n_0t)-\sqrt{n+2t}\right) \ . \qquad\qquad
\label{B-10b}
\end{eqnarray}
\end{subequations}
After a simple calculation, $F_1$ is reduced to 
\begin{eqnarray}
& &F_1=\sqrt{n_0+2t}\left[1-\Delta_0(n_0,t)\right] e^{n_0} \ , 
\label{B-11}\\
& &\Delta_0(n_0,t)=(1/2)n_0\cdot (n_0+2t)^{-1}
\left[\sqrt{n_0+2t}+\sqrt{n_0+1+2t}\right]^{-2} \ . 
\label{B-12}
\end{eqnarray}
Here, we used the relation 
\begin{equation}\label{B-13}
\sum_{n=0}^{\infty}\frac{n_0^n}{n!}(n-n_0)^2 = n_0 e^{n_0} \ . 
\end{equation}
The part $F_2$ is expressed in the form 
\begin{eqnarray}
& &F_2=-\sqrt{n_0+2t}\cdot \Delta_0(n_0,t)/n_0 \nonumber\\
& &\qquad\quad \times \sum_{n=0}^{n_0}\frac{n_0^n}{n!}
(n_0-n)^2(n_0-n+1)\cdot g(n;n_0t) \ , 
\label{B-14}\\
& &g(n;n_0t)=\frac{\sqrt{n+2t}+\sqrt{n_0+1+2t}+2\sqrt{n_0+2t}}
{(\sqrt{n+2t}+\sqrt{n_0+2t})^2(\sqrt{n+2t}+\sqrt{n_0+1+2t})^2} \ . 
\label{B-15}
\end{eqnarray}
Since $g(x;n_0t)>0$, $g'(x;n_0t)<0$ and 
$g''(x;n_0t)>0$ in the region $0 \leq x \leq n_0$, 
$g(n;n_0t)$ satisfies 
\begin{equation}\label{B-16}
g(n;n_0t) \leq g(n_0;n_0t)+(g(n_0;n_0t)-g(0;n_0t))/n_0\cdot
(n-n_0) \ . 
\end{equation}
Therefore, $F_2$ obeys the following inequality : 
\begin{eqnarray}\label{B-17}
F_2 > &-&\sqrt{n_0+2t}\cdot \Delta_0(n_0;t)/n_0\cdot 
\biggl[ g(n_0;n_0t)\sum_{n=0}^{n_0}\frac{n_0^n}{n!}(n_0-n)^2(n_0-n+1) 
\nonumber\\
&+& (g(0;n_0t)-g(n_0;n_0t))/n_0\cdot \sum_{n=0}^{n_0}
\frac{n_0^n}{n!}(n_0-n)^3(n_0-n+1)\biggl] \ . 
\end{eqnarray}
%For the derivation of the relation (\ref{B-7}), 
In the succeeding calculation, 
the following formula is useful : 
\begin{eqnarray}\label{B-18}
& &\sum_{n=0}^{n_0}\frac{n_0^n}{n!}(n_0-n)^2
=n_0\sum_{n=0}^{n_0}\frac{n_0^n}{n!}-n_0\cdot \frac{n_0^{n_0}}{n_0!} \ , 
\nonumber\\
& &\sum_{n=0}^{n_0}\frac{n_0^n}{n!}(n_0-n)^3
=n_0(2n_0+1)\frac{n_0^{n_0}}{n_0!}
-n_0\sum_{n=0}^{n_0}\frac{n_0^n}{n!} \ , 
\nonumber\\
& &\sum_{n=0}^{n_0}\frac{n_0^n}{n!}(n_0-n)^4
=n_0(3n_0+1)\sum_{n=0}^{n_0}\frac{n_0^n}{n!}-n_0(6n_0+1)\cdot 
\frac{n_0^{n_0}}{n_0!} \ .
\end{eqnarray}
The relation (\ref{B-18}) leads us to 
\begin{eqnarray}\label{B-19}
\sum_{n=0}^{n_0}\frac{n_0^n}{n!}(n_0-n)^2(n_0-n+1)
&=&2n_0^2\frac{n_0^{n_0}}{n_0!} \nonumber\\
&\simeq& 2n_0^2 D(n_0)e^{n_0} \ , \nonumber\\
\sum_{n=0}^{n_0}\frac{n_0^n}{n!}(n_0-n)^3(n_0-n+1)
&=&3n_0^2\sum_{n=0}^{n_0}\frac{n_0^n}{n!}-4n_0^2\frac{n_0^{n_0}}{n_0!} 
\nonumber\\
&\simeq& n_0^2 [3C(n_0)-4D(n_0)]e^{n_0} \ .
\end{eqnarray}
Then, the inequality (\ref{B-17}) becomes 
\begin{eqnarray}\label{B-20}
F_2 > &-&\sqrt{n_0+2t}\cdot \Delta_0(n_0;t) \nonumber\\
&\times& \biggl[2((n_0+2)g(n_0;n_0t)-2g(0;n_0t))D(n_0) \nonumber\\
& &\ \ +3(g(0;n_0t)-g(n_0;n_0t)) C(n_0) \biggl] e^{n_0}\ . 
\end{eqnarray}
The relations (\ref{B-10}), (\ref{B-12}) and (\ref{B-20}) give us 
\begin{equation}\label{B-21}
\sum_{n=0}^{\infty}\frac{n_0^n}{n!} \sqrt{n+2t}\cdot e^{-n_0} 
> \sqrt{n_0+2t}\ (1-\Delta(n_0,t)) \ , 
\end{equation}
\vspace{-0.3cm}
\setcounter{equation}{20}
\begin{subequations}\label{B-21a}
\begin{eqnarray}
\Delta(n_0,t)&=&\Delta_0(n_0,t) \nonumber\\
& &\times \biggl[ 1+2((n_0+2)g(n_0;n_0t)-2g(0;n_0t))D(n_0) \nonumber\\
& &\qquad +3(g(0;n_0t)-g(n_0;n_0t))C(n_0) \biggl] \ . 
\end{eqnarray}
\end{subequations}
On the other hand, we have the relation 
\begin{eqnarray}\label{B-22}
& &\sum_{n=0}^{\infty} \frac{n_0^n}{n!} \left( \sqrt{n_0+2t}-\sqrt{n+2t}
\right) \nonumber\\
&=&(1/2)\left(\sqrt{n_0+2t}\right)^{-1}
\sum_{n=0}^{\infty}\frac{n_0^n}{n!}\left(\sqrt{n_0+2t}-
\sqrt{n+2t}\right)^2 \geq 0 \ , 
\end{eqnarray}
namely, 
\begin{equation}\label{B-23}
\sqrt{n_0+2t} \geq \sum_{n=0}^{\infty}\frac{n_0^n}{n!}\sqrt{n+2t}\cdot
e^{-n_0} \ . 
\end{equation}

%%%%%%%%%%%%%%%%%%%%%%%%%%%%% (Fig.1) %%%%%%%%%%%%%%%%%%%%%%%%%%%%%%%
\begin{figure}[t]
  \epsfxsize=10cm  % \epsfysize=   cm  
  \centerline{\epsfbox{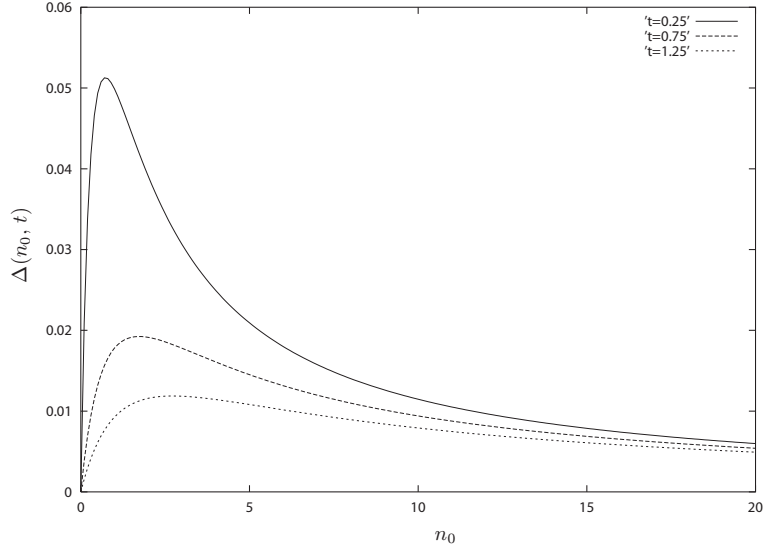}}
  \caption{$\Delta(n_0,t)$ with $t=1/4, 3/4$ and $5/4$ are depicted.}
   \label{fig:1}
\end{figure}
%%%%%%%%%%%%%%%%%%%%%%%%%%%%%%%%%%%%%%%%%%%%%%%%%%%%%%%%%%%%%%%%%%%%

Thus, the relations (\ref{B-21}) and (\ref{B-23}) give us 
\begin{equation}\label{B-24}
\sqrt{n_0+2t} \geq \sum_{n=0}^{\infty}\frac{n_0^n}{n!}\sqrt{n+2t}\cdot
e^{-n_0} > \sqrt{n_0+2t}\ (1-\Delta(n_0,t)) \ . 
\end{equation}
The above inequality is derived under the condition 
$n_0=0,1,2,\cdots$. 
We regard this relation as that satisfying in the case of any positive number. 
%Tables IIIa, IIIb and IIIc 
Figure 1 shows numerical results of $\Delta(n_0,t)$ 
and, through these results, we can conclude that the relation 
(\ref{7-8}) may be acceptable.


\begin{thebibliography}{99}
%%%%%%%%%%%%%%%%%%%%%%%%%%%%%%%%%%%%%%%%%%%%%%%%%%%%%%%%%%%%%
% Some macros are available for the bibliography:
%   o for general use
%      \JL : general journals          \andvol : Vol (Year) Page
%   o for individual journal 
%      \PR  : Phys. Rev.               \PRL : Phys. Rev. Lett.
%      \NP  : Nucl. Phys.              \PL  : Phys. Lett.
%      \JMP : J. Math. Phys.           \CMP : Commun. Math. Phys.
%      \PTP : Prog. Theor. Phys.       \JPSJ: J. Phys. Soc. Jpn.
%      \JP  : J. of Phys.              \NC  : Nouvo Cim.
%      \IJMP: Int. J. Mod. Phys.       \ANN : Ann. of Phys.
% Usage:
%   \PR{D45,1990,345}            ==> Phys.~Rev.\ {\bf D45} (1990), 345
%   \JL{Phys.~Lett.,A30,1981,56} ==> Phys.~Lett.\ {\bf A30} (1981), 56
%   \andvol{B123,1995,1020}      ==> {\bf B123} (1995), 1020
%%%%%%%%%%%%%%%%%%%%%%%%%%%%%%%%%%%%%%%%%%%%%%%%%%%%%%%%%%%%%
\bibitem{one}
A. Kuriyama, C. Provid\^encia, J. da Provid\^encia, Y. Tsue and M.~Yamamura, 
in preparation. 
\bibitem{two}
A. Kuriyama, C. Provid\^encia, J. da Provid\^encia, Y. Tsue and M.~Yamamura, 
        Prog. Theor. Phys. {\bf 106} (2001), 751. 
\bibitem{three}
A. Kuriyama, C. Provid\^encia, J. da Provid\^encia, Y. Tsue and M.~Yamamura, 
        Prog. Theor. Phys. {\bf 107} (2002), 1273.
\bibitem{four}
A. Kuriyama, C. Provid\^encia, J. da Provid\^encia, Y. Tsue and M.~Yamamura, 
        Prog. Theor. Phys. {\bf 107} (2002), 1279. 
\bibitem{five}
A. Kuriyama, C. Provid\^encia, J. da Provid\^encia, Y. Tsue and M.~Yamamura, 
        Prog. Theor. Phys. {\bf 106} (2001), 765, ibid. {\bf 107} (2002), 65, 
        ibid. {\bf 107} (2002), 443, ibid. {\bf 108} (2002), 323, 
        ibid. {\bf 108} (2002), 783, ibid. {\bf 109} (2003), 77. 
\bibitem{six}
K. A. Penson and A. I. Solomon, 
        J. Math. Phys. {\bf 40} (1999), 2354. 
\bibitem{seven}
T. Marumori, M. Yamamura and A. Tokunaga, Prog. Theor. Phys. {\bf 31} (1964), 
1009. 
\bibitem{eight}
T. Marumori, T. Maskawa, F. Sakata and A. Kuriyama, Prog. Theor. Phys. 
{\bf 64} (1980), 1294. \\
M. Yamamura and A> Kuriyama, Prog. Theor. Phys. Suppl. No. 93 (1987), 1. 
\bibitem{nine}
A. Kuriyama, C. Provid\^encia, J. da Provid\^encia, Y. Tsue and M.~Yamamura, 
        Prog. Theor. Phys. {\bf 104} (2000), 155. 
\end{thebibliography}
\end{document}